\documentclass[preprint,12pt,authoryear]{elsarticle}

\usepackage{amssymb, amsmath, amsthm, extarrows, mathtools, bm, esvect}
\usepackage{color, graphicx, tikz}
\usepackage[font=small]{caption}
\usepackage{subcaption, float}
\usepackage{multicol, multirow}
\usepackage{enumitem, verbatim}
\usepackage{hyperref}
\usepackage{adjustbox}
\usepackage{booktabs}
\usepackage{rotating}

\usepackage{tikz}
\usetikzlibrary{spy}

\newcommand{\R}{\mathbb{R}}

\newlength{\spyimagewidth}
\newlength{\spyimageheight}

\DeclareMathOperator{\DB}{DB}
\DeclareMathOperator{\Uref}{\textit{u}^{\text{Ref}}}

\newcommand{\spyPalma}[2][0.96\linewidth]{%
  \setlength{\spyimagewidth}{#1}
  \setlength{\spyimageheight}{\spyimagewidth}
  \begin{tikzpicture}[spy using outlines={rectangle, magnification=2.9, size=0.5\spyimagewidth, connect spies,color=white}]
\node[inner sep=0pt, anchor=south west] (image) at (0,0) {\adjustbox{trim=17.5cm 32cm 21.5cm 7cm,clip,width=\spyimagewidth}{\includegraphics{#2}}};
    \coordinate (spy point) at (0.82\spyimagewidth , 0.47\spyimageheight);
    \coordinate (spy node) at (0.25\spyimagewidth, 0.6\spyimageheight);
    \spy on (spy point) in node at (spy node);
  \end{tikzpicture}%
}

\newcommand{\spyPoland}[2][0.96\linewidth]{%
  \setlength{\spyimagewidth}{#1}
  \setlength{\spyimageheight}{\spyimagewidth}
  \begin{tikzpicture}[spy using outlines={rectangle, magnification=2.5, size=0.4\spyimagewidth, connect spies,color=white}]
    \node[inner sep=0pt, anchor=south west] (image) at (0,0) {\adjustbox{trim=0cm 0cm 0cm 0cm,clip,width=\spyimagewidth}{\includegraphics{#2}}};
    \coordinate (spy point) at (0.40\spyimagewidth , 0.6\spyimagewidth );
    \coordinate (spy node) at (0.2\spyimagewidth, 0.2\spyimageheight);
    \spy on (spy point) in node at (spy node);
  \end{tikzpicture}%
}

\newcommand{\spyBCN}[2][0.96\linewidth]{%
  \setlength{\spyimagewidth}{#1}
  \setlength{\spyimageheight}{\spyimagewidth}
  \begin{tikzpicture}[spy using outlines={rectangle, magnification=2.9, size=0.5\spyimagewidth, connect spies,color=white}]
    \node[inner sep=0pt, anchor=south west] (image) at (0,0) {\adjustbox{trim=0cm 0cm 0cm 0cm,clip,width=\spyimagewidth}{\includegraphics{#2}}};
    \coordinate (spy point) at (0.63\spyimagewidth , 0.39\spyimageheight);
    \coordinate (spy node) at (0.25\spyimagewidth, 0.25\spyimageheight);
    \spy on (spy point) in node at (spy node);
  \end{tikzpicture}%
}

\newcommand{\spyKorea}[2][0.96\linewidth]{%
  \setlength{\spyimagewidth}{#1}
  \setlength{\spyimageheight}{\spyimagewidth}
  \begin{tikzpicture}[spy using outlines={rectangle, magnification=1.8, size=0.4\spyimagewidth, connect spies,color=white}]
    \node[inner sep=0pt, anchor=south west] (image) at (0,0) {\adjustbox{trim=0cm 0cm 0cm 0cm,clip,width=\spyimagewidth}{\includegraphics{#2}}};
    \coordinate (spy point) at (0.69\spyimagewidth , 0.77\spyimagewidth );
    \coordinate (spy node) at (0.2\spyimagewidth, 0.2\spyimageheight);
    \spy on (spy point) in node at (spy node);
  \end{tikzpicture}%
}

\journal{ISPRS Journal of Photogrammetry and Remote Sensing}

\begin{document}

\begin{frontmatter}



\title{Super-Resolution of Sentinel-2 Images Using a Geometry-Guided Back-Projection Network with Self-Attention}


\author{I.~Pereira-Sánchez, D.~Torres, F.~Alcover, B.~Garau, J.~Navarro, C.~Sbert, J.~Duran} 

\affiliation{organization={Dpt.~Mathematics and Computer Science and IAC3, University of the Balearic Islands (UIB)},
            addressline={Cra.~de Valldemossa, km.~7.5}, 
            city={Palma},
            postcode={E-07122}, 
            state={Illes Balears},
            country={Spain}}

\begin{abstract}

The Sentinel-2 mission provides multispectral imagery with 13 bands at resolutions of 10m, 20m, and 60m. In particular, the 10m bands offer fine structural detail, while the 20m bands capture richer spectral information. In this paper, we propose a geometry-guided super-resolution model for fusing the 10m and 20m bands. Our approach introduces a cluster-based learning procedure to generate a geometry-rich guiding image from the 10m bands. This image is integrated into an unfolded back-projection architecture that leverages image self-similarities through a multi-head attention mechanism, which models nonlocal patch-based interactions across spatial and spectral dimensions. We also generate a dataset for evaluation, comprising three testing sets that include urban, rural, and coastal landscapes. Experimental results demonstrate that our method outperforms both classical and deep learning-based super-resolution and fusion techniques.
\end{abstract}


\end{frontmatter}


\section{Introduction}

The Sentinel-2 (S2) mission is an Earth observation initiative encompassed in the \href{https://www.copernicus.eu/en}{Copernicus program} of the European Space Agency (ESA). It consists of two identical satellites\footnote{At the time of writing, satellite 2A, which was replaced by 2C in January 2025, has been performing an extension campaign since March 2025, exceptionally complementing the two other satellites.}, 2B and 2C, orbiting with a $180^{\circ}$ phase delay in order to offer a 5-day revisit cycle at the equator \citep{sentinel2-wiki, 2012-malenovsky}. These frequent revisits, combined with global coverage and an open-access policy, are among the key reasons why the S2 mission plays a crucial role in remote sensing applications. S2 imagery has been used for a wide range of tasks such as bathymetry \citep{2024Bathymetry}, crop and vegetation monitoring \citep{2016-vegetation, 2020-agriculture}, studying water bodies and the marine ecosystem \citep{2016-water, 2018-seagrass}, or evaluating the impact of human-induced changes in the landscape \citep{2017-construction}. For these applications to be effective, high-quality imagery is essential to capture (almost) real-time, dynamic changes on a macroscopic scale. Due to constraints of storage, transmission, and hardware limitations, the bands are acquired at different levels of spatial resolution.

\begin{figure}[t]
    \centering
    \includegraphics[width=0.75\linewidth]{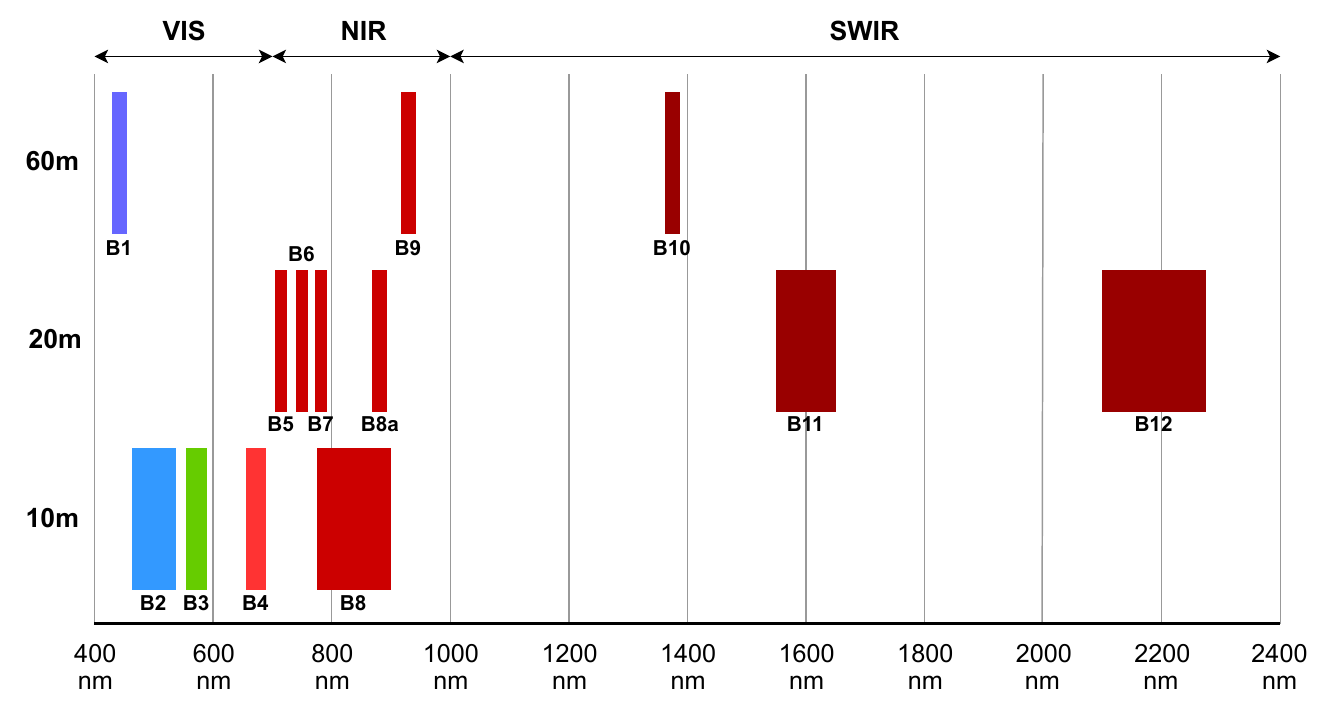}
    \caption{Sentinel-2 satellites provide four spectral bands at 10m resolution covering the visible and near-infrared (NIR) spectrum, six at 20m resolution in the NIR and short-wave infrared (SWIR) ranges, and three additionals bands at 60m, commonly used for atmospheric correction.}
    \label{fig:dibuixbandes}
\end{figure}

S2 satellites provide 13 spectral bands, as illustrated in Figure \ref{fig:dibuixbandes}. These include four bands at 10m resolution covering the visible and near-infrared (NIR) spectrum, six bands at 20m resolution, some of which fall within the short-wave infrared (SWIR) range, and three additional bands at 60m resolution \citep{2016-water}. While the coarser bands offer less detailed spatial representation of ground features, they provide unique spectral information that can be valuable in specific applications.  For instance, the 20m bands are particularly useful for vegetation analysis, while the 60m bands contribute to pre-processing tasks such as atmospheric correction \citep{2023-wu-hyerarchical}. Ideally, having access to the full spectral range at high spatial resolution would significantly enhance the accuracy and effectiveness of downstream tasks.

Improving the spatial resolution of S2 data involves addressing the inherent differences in scale among its spectral bands. A straightforward approach is to apply general upsampling methods, such as bicubic interpolation or multi-temporal band analysis \citep{tarasiewicz2023multitemporal}, to each low-resolution (LR) image. However, satellite imagery enables more refined strategies by guiding the super-resolution (SR) process with the high-resolution (HR) bands. Many satellites provide a HR panchromatic (PAN) image that covers a broad range of wavelengths. When such PAN data are available, the fusion problem is referred to as \textit{pansharpening} if the LR image contains a few bands, or as \textit{hypersharpening} if it contains dozens or even hundreds of spectral bands. This problem has been widely studied in the literature \citep{pereira2024comprehensive}. Among classical methods, component substitution \citep{gillespie1987color,shettigara1992generalized}, multi-resolution analysis \citep{aiazzi2003mtf,khan2008indusion}, and variational models \citep{ballester2006variational,duran2014nonlocal} have been mainly proposed. In the deep learning domain, some networks have adopted residual architectures \citep{cai2020super,Yang_2017_ICCV} and attention mechanisms \citep{lu2023awfln}, including transformer-based designs \citep{zhou2022panformer}. More recently, deep unfolding methods \citep{mai2024deep, zhang2023spatial}, which combine the strengths of model-based and purely data-driven deep learning approaches, have emerged.

Sentinel-2 does not include a PAN band, which complicates the direct application of fusion methods. To overcome this limitation, some approaches simulate a PAN-like image by leveraging the HR bands. Two strategies are commonly used: selecting a single HR band for each coarse one based on spectral similarity \citep{gavsparovic2018effect,wang2015newgeostatistical}, or constructing a PAN image as a linear combination of all HR bands \citep{selva2015hypersharpening,wang2016}. While this enables the use of fusion algorithms, the synthetic PAN image does not fully capture the spectral diversity of the coarse bands, limiting the quality of the results. Alternative techniques directly guide the upsampling process using the available HR data, mainly through deep learning tools \citep{2018-lanaras-network,lim2017edsr,2023-wu-hyerarchical}, although some classical methods based on variational formulations have also been proposed \citep{lanaras2017super,2019ReducedRank,2020SSSS}.

In this paper, we address the problem of integrating the geometric accuracy of the 10m bands with the spectral consistency of the 20m ones. To achieve this, we introduce a cluster-based learnable module, adapted from \citep{2025SSSR}, which generates a guiding image with six channels (one per 20m band). This module first clusters the pixels of the 10m bands and then maps each cluster into the spectral space of the 20m bands using a Multi-Layer Perceptron (MLP) with cluster-specific weights. We compare this approach with the similarity-based strategy introduced in \citep{gavsparovic2018effect, S2PanCompar2018}, which selects as guiding image the 10m band closest in wavelength to the target 20m band. On the other hand, we propose a geometry-guided SR model based on an unfolded back-projection method that minimizes the error defined by the observation model for LR data. With the aim of evading the artifacts which are known to appear during the reconstruction process \citep{dai2007bilateral}, the back-projection kernel is replaced by a residual network that incorporates a nonlocal module. This module leverages the self-similarity in satellite images and is built upon the Multi-Head Attention (MHA) mechanism \citep{dosovitskiy2020image}, which integrates the geometry detail from the guiding image.

Our main contributions can be summarized as follows:
\begin{itemize}
    \item We propose an unfolded back-projection method built upon the classical observation model for LR data and a nonlocal residual architecture that exploits image self-similarities through MHA. The attention heads are designed to capture spatial and spectral information by computing similarity weights across the back-projected error and a guiding image constructed from the 10m bands. The similarity is measured in terms of patches rather than individual pixels. To the best of our knowledge, there are no works using back-projection to enhance the resolution of Sentinel-2 data.

    \item We introduce a cluster-based learning procedure to derive geometry-rich guiding images from the 10m bands needed to enhance the spatial resolution of the 20m bands.

    \item We generate a dataset for Sentinel-2 super-resolution using Wald's protocol~\citep{wald1997fusion}. The testing set consists of urban, rural, and coastal areas, allowing us to evaluate performance on different scenes commonly studied in remote sensing applications.
\end{itemize}

The rest of the paper is organized as follows. Section \ref{sec:RelatedWork} reviews related work in SR and image fusion within the Sentinel-2 context. In Section \ref{sec:Model}, we introduce the proposed unfolded back-projection model and the learning procedure used to build the guiding image. Section \ref{sec:dataset} describes the proposed dataset, and discusses several composites and indexes involving the 20m bands. An extensive performance comparison is presented in Section \ref{sec:ExpResults}, while Section \ref{sec:Ablation} conducts an ablation study to assess the contribution of different components of the model and highlight the optimal configurations. Finally, conclusions are drawn in Section \ref{sec:Conclusion}.

\section{Related work} \label{sec:RelatedWork}

There are several approaches to increasing the spatial resolution of Sentinel-2 images. The simplest strategy involves operating on a single LR band using either classical or deep learning-based models. This has been particularly investigated in the visible and NIR wavelengths. In this setting, Liebel et~al.~\citep{liebel2016single} applied the CNN proposed by Dong et~al. \citep{dong2014learning} for SR to enhance the RGB channels, training on downsampled images while treating the 10m bands as reference. Galar et al.~\citep{galar2019super} adopted the deep residual network introduced by Lim et~al \citep{lim2017edsr} to enhance the VNIR bands, using RapidEye imagery as reference. More recently,  Aldo{\u{g}}an et al.~\citep{aldougan2024enhancement} increased the resolution of S2 imagery with a GAN pre-trained on PlanetScope data to obtain enhanced images for ship detection.

Another approach is multi-temporal image SR, which exploits the 5-day revisit cycle of the S2 satellites to gather additional information across multiple time frames. Vaquerio et~al.~\citep{vaqueiro2021multi} proposed an algorithm that leverages both the revisit cycle and the georeferencing error between images to improve spatial resolution using data from the $K$ nearest pixels. This idea has been further developed and integrated into deep learning architectures~\citep{deudon2020highres, razzak2023multi}.

\subsection{Guided super-resolution of the Sentinel-2 coarse bands}

Guided super-resolution, which is conceptually equivalent to image fusion, seeks to enhance the spatial resolution of the 20m/60m bands by leveraging the geometric accuracy of the 10m bands. The HR images serve a reference to guide the reconstruction of spatial details that are missing or blurred in the coarser bands. Such guidance can be exploited through explicit spatial priors, edge-preserving regularization, self-similarity, or learned attention mechanisms.

In the variational setting, Lanaras et al.~\citep{lanaras2017super} proposed a convex energy functional that integrates the classical observation model for LR images within a reduced-dimensional subspace, combined with adaptive edge-reserving regularization that promotes self-similarity guided by the 10m bands. The subspace is learned from the input data to reduce the number of unknowns. Although not specifically designed for S2 imagery, this approach was later evaluated in \citep{zhang2019super}. 
Ulfarsson et al.~\citep{2019ReducedRank} introduced a similar formulation, although the low-dimensional subspace is not fixed but automatically estimated during optimization, resulting in a nonconvex problem. The solution is approximated using a conjugate gradient and trust-region based cyclic descent algorithm. Building on the two previous works, Lin and Bioucas-Dias~\citep{2020SSSS} defined a convex, scene-adapted regularizer based on a self-similarity graph derived from the 10m bands.

Deep learning models have become widely used for enhancing S2 data. Residual networks for SR were adapted to the S2 setting by Palsson et al.~\citep{ResNet2018} and Lanaras et al.~\citep{2018-lanaras-network}. These supervised models require reference images for training, which are not readily available. To address this, some approaches assume the mappings $20\to 10$m and $40\to20$m to be roughly equivalent, learning the latter to then upsample the 20m bands to 10m resolution. Wu et al.~\citep{2023-wu-hyerarchical} introduced these ideas into a hierarchical network that exploits self-similarity among the coarser bands before the fusion step. Wang et al.~\citep{2019WangSVR} used a Support Vector Regression (SVR) model to find the function that minimizes the error between the prediction and the target, while allowing for small deviations. Other architectures rely on GANs \citep{salgueiro2020super}.

The lack of reference images to compare the resulting sharpened bands motivates the use of unsupervised networks. Nguyen et al.~\citep{nguyen2021sentinel} proposed a loss function based on the observation model for LR images, incorporating the Modulation Transfer Functions (MTF) of the sensors. Other approaches, such as Fern\'andez et al.~\citep{fernandez2018multimodal}, employ multimodal probabilistic latent semantic analysis. Among unsupervised models, the unfolded network by Nguyen et al.~\citep{2023Nguyen} stands out as an example of combining traditional and learning-based~strategies.

The use of complementary information from S2 and other satellites has also been explored. For instance, Kremezi et al.~\citep{worldview-sentinel2} combine the high spatial resolution of WorldView-3 imagery with the spectral accuracy of S2 data using various fusion methods for marine plastic litter monitoring. More recently, Alparone et al.~\citep{SpatialResEnhanc2024} iteratively fuses S2 and PRISMA data up to a 5m resolution, using the previously enhanced images as guidance for the subsequent fused steps.

\subsection{Guiding image selection from Sentinel-2 fine bands} \label{PANSOTASubsection}

A crucial component of a guided SR model is the selection of an appropriate guiding image, which captures the geometry from the 10m bands to be incorporated during the upsampling of the 20m bands. In some literature, these guiding images are referred to as panchromatic in the context of Sentinel-2. However, this terminology is avoided here, as they do not contain information spanning the entire wavelength spectrum.

The straightforward approach presented in \citep{gavsparovic2018effect} consists of selecting the guiding image based on the spectral similarity or distance to the target coarse band. As illustrated in Figure \ref{fig:dibuixbandes}, bands B8a, B11, and B12 are associated with B8, as it is closer in wavelength. However, for B5, B6, and B7, the average of B4 and B8 is used. Spectral similarity can also be studied in terms of the correlation coefficient \citep{wang2015newgeostatistical}. In the S2 setting, this measures the correlation between downsampled 10m bands and the 20m bands \citep{2016-water}.

Selecting a specific HR band as the guiding image, while straightforward, may not adequately capture the complex relationships among all bands. In fusion models, the PAN image generally encompasses information across a broad wavelength range and is commonly represented as a linear combination of the spectral bands. Selva et al.~\citep{selva2015hypersharpening} approximate this combination by computing the simple average of all 10m bands. However, this approach may inadequately represent the true relationships between spectral channels, potentially incorporating information unrelated to the 20m bands in S2 data. A more robust alternative involves employing a weighted average, where the weights are determined through multiple regression, as proposed by Wang et al.~\citep{wang2016}. This methodology enables the synthesized guiding images to capture more intricate inter-band dependencies compared to methods relying solely on the most correlated band. The correlation-based scheme presented by Park et al.~\citep{2017ModifiedSelSynthBand} offers another strategy for deriving the weights in this linear combination.

\subsection{Back-Projection algorithms}

Back-projection algorithms \citep{Irani1991BackProj} are an efficient image refinement iterative procedure, where at each iteration the reconstruction error is back-projected into the target image to improve spatial details. Originally developed for tomographic reconstruction \citep{peters1981algorithms,medoff1983iterative}, these techniques were later adapted to address other tasks, such as image SR~\citep{irani1993motion,zomet2001robust}.

In recent years, back-projection has been adapted to deep learning by replacing the classical operators involved in each iteration with learnable neural network modules. For instance, Haris et~al.~\citep{haris2018deep} introduced an unfolded back-projection network that jointly learns up- and down-projection units using convolutional layers. Their model considers various types of image degradation and captures HR components by deeply concatenating features across multiple up- and downsampling steps. More recently, Zhang et al.~\citep{zhang2023spatial} introduced an unfolded back-projection network for pansharpening, which comprises two back-projection schemes unfolded in parallel: one mapping the LR spectral image into the spatial domain, and the other projecting the PAN image into the spectral domain. The final fused product is obtained by combining the spatial and spectral features produced by both schemes.

\section{Proposed model} \label{sec:Model}

In this section, we present the proposed geometry-guided unfolded back-projection SR model for upsampling the 20m bands of Sentinel-2 imagery to 10m resolution. We also address the selection of the guiding image, comparing the classical spectral similarity-based strategy \citep{gavsparovic2018effect} with the approach proposed in this work, which relies on a cluster-based learning procedure. We refer to the model employing spectral similarity-based guidance as GINet, and to the one using cluster-based guidance as GINet+.

\subsection{Geometry-guided super-resolution method}

Let $f \in \R^{C\times M}$ denote the input data containing the LR bands as channels, and let $u\in\R^{C\times N}$ be the corresponding enhanced image to be estimated. In the Sentinel-2 setting, $C=6$, as this is the number of 20m bands, $N$ represents the number of pixels in the 10m resolution domain, $M=\tfrac{N}{s^2}$, and the sampling factor is $s=2$. 

The first objective is to increase the spatial resolution of $f$ while keeping its spectral information. To do this, we base our formulation on the classical observation model for SR:
\begin{equation} \label{LowResModel}
    f = \DB u + \eta,
\end{equation}
where $\text{B}:\mathbb{R}^{C\times N}\to\mathbb{R}^{C\times N}$ is a low-pass filter, $\text{D}:\mathbb{R}^{C\times N}\to \mathbb{R}^{C\times M}$ is an $s$-fold decimation, and $\eta$ is the realization of noise.

To recover $u$, back-projection methods iterate 
$$
u^{k+1}=u^k + \mathcal{K} \ast (\DB u^k-f)_{\uparrow},
$$
where $\uparrow$ represents bicubic interpolation and $\mathcal{K}$ is a back-projection kernel that controls the speed of convergence. However, as back-projecting a reconstruction error can produce artifacts caused by the ringing effect \citep{dai2007bilateral}, especially around contrasted edges, a nonlocal filter is often added to the model \citep{dong2009nonlocal}. Accordingly, we propose the following nonlocal (NL) back-projection scheme:
\begin{equation} \label{NLBPDetailInjection}
\begin{cases}
     e^{k} =  \left(\DB u^k-f\right)_{\uparrow},  \\[0.5ex]
    u^{k+1} = u^k + \mathcal{K} \ast \mathrm{NL}(e^{k}).
\end{cases}
\end{equation}

Once this scheme has been established, we unfold it and replace the involved operators with learning-based networks. Classical optimization typically require hundreds of iterations. However, due to the computational demands of training, we rely on a reduced number. To avoid confusion with training epochs, these optimization steps will be referred to as {\it stages}. Specifically, we replace $\DB$ with the composition of a 2D convolution and average pooling, and $\uparrow$ with a transposed convolution. In the first stage, we set $u^0$ as the bicubic interpolation of $f$.

\begin{figure}[t]
  \centering
  \includegraphics[trim=1.5cm 1cm 1.5cm 0cm, clip=true, width=0.76\linewidth]{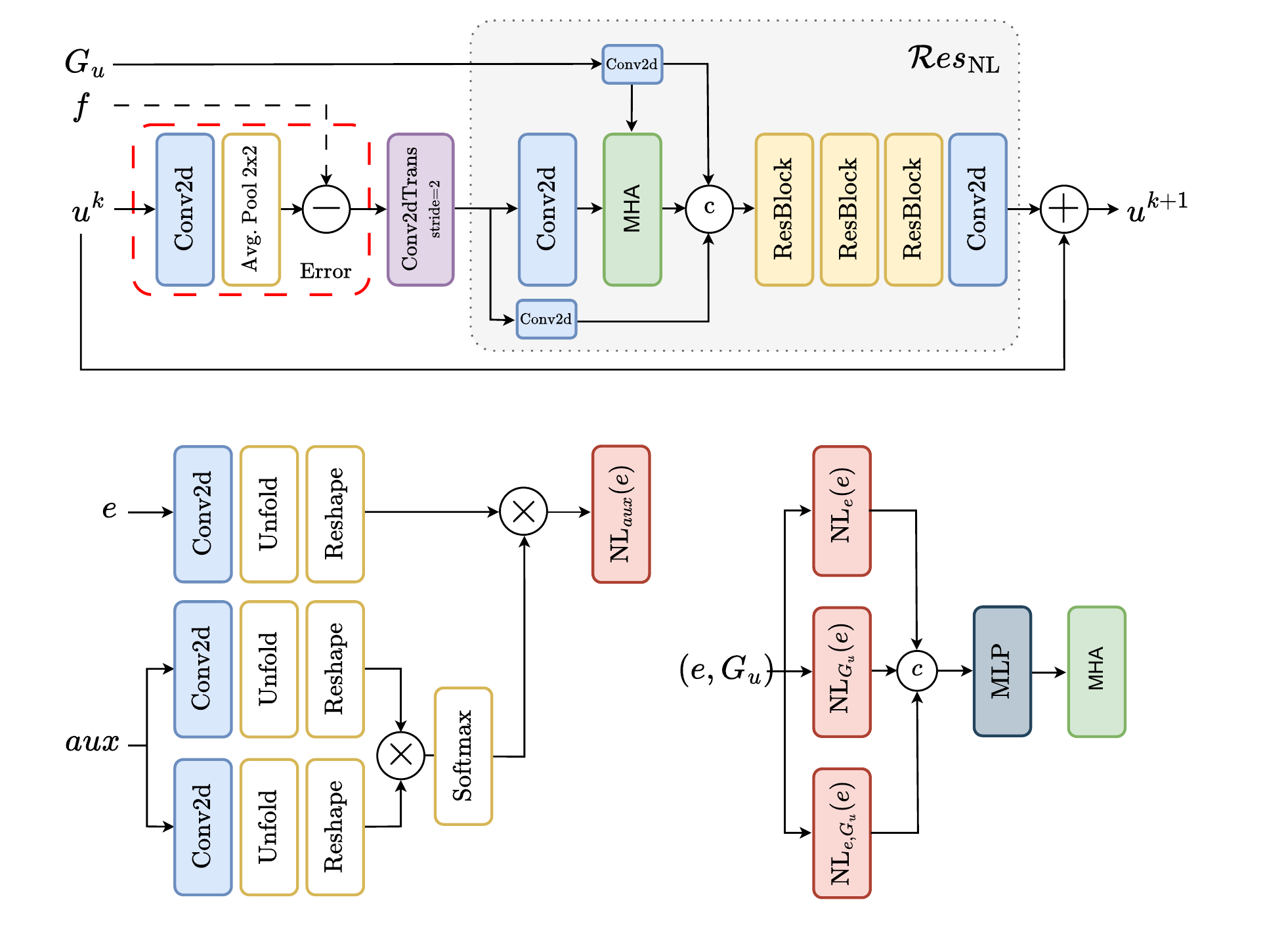}
  \caption{Architecture of the proposed method. The top diagram illustrates one stage, where $G_u$ denotes the guiding image. The error is computed as defined in \eqref{NLBPDetailInjection}, where $\DB$ has been replaced by a convolution followed by average pooling. Then, it is upsampled by a transposed convolution. Each ResBlock in $\mathcal{R}es_{\mathrm{NL}}$ consists of two convolutional layers followed by a residual connection. The architecture of the MHA module is shown in the bottom-right, where three heads are considered in parallel using the upsampled error map, the guiding image, and their concatenation as inputs. Each head, illustrated in the bottom-left, computes similarity weights based on patches extracted from the corresponding auxiliary image, while the nonlocal filtering is uniformly applied to the upsampled error map across all heads.}
  \label{fig:architecture}
\end{figure}

The convergence kernel $\mathcal{K}\ast \text{NL}$ in the back-projection scheme is replaced with $\mathcal{R}es_{\mathrm{NL}}$, a residual network inspired by Pereira-Sánchez et~al. \citep{MARNet} that integrates a MHA module with three attention heads operating in parallel. The residual part consists of three ResBlocks, each composed of two convolutional layers followed by a residual connection. The MHA computes the similarity weights independently for each head, respectively using the upsampled error, the guiding image $G_u$, and their concatenation as reference. In contrast, the nonlocal filter is consistently applied to the upsampled error across all heads. We propose to compute the similiarity weights as
\begin{equation*}
\omega_{i,j} = \dfrac{1}{\Gamma_i} \exp\left(\theta(g_{Q_i})^t \phi(g_{Q_j}) \right),
\end{equation*}
where $g$ denotes any reference image, $\Gamma_i$ is a normalization factor, $\theta$ and $\phi$ are learnable convolutional filters, and $Q_i$ and $Q_j$ are patches centered at pixels $i$ and $j$, respectively. Additionally, to improve computational efficiency, interactions between neighbouring pixels are restricted to predefined windows instead of the entire image. 

Figure \ref{fig:architecture} illustrates the architecture of the proposed model. The learnable parameters of the modules are not shared across stages.

\subsection{Guiding image generation} \label{SubsectionPANGen}

We compare two distinct strategies for computing the guiding image by extracting the geometric information encoded in the 10m Sentinel-2 bands. The first follows the approach proposed in \citep{gavsparovic2018effect}, which consists of selecting as guiding image the 10m band closest in wavelength to the target 20m band (refer back to Subsection \ref{PANSOTASubsection}). 

\begin{figure}[t]
  \centering
  \includegraphics[width=0.85\linewidth]{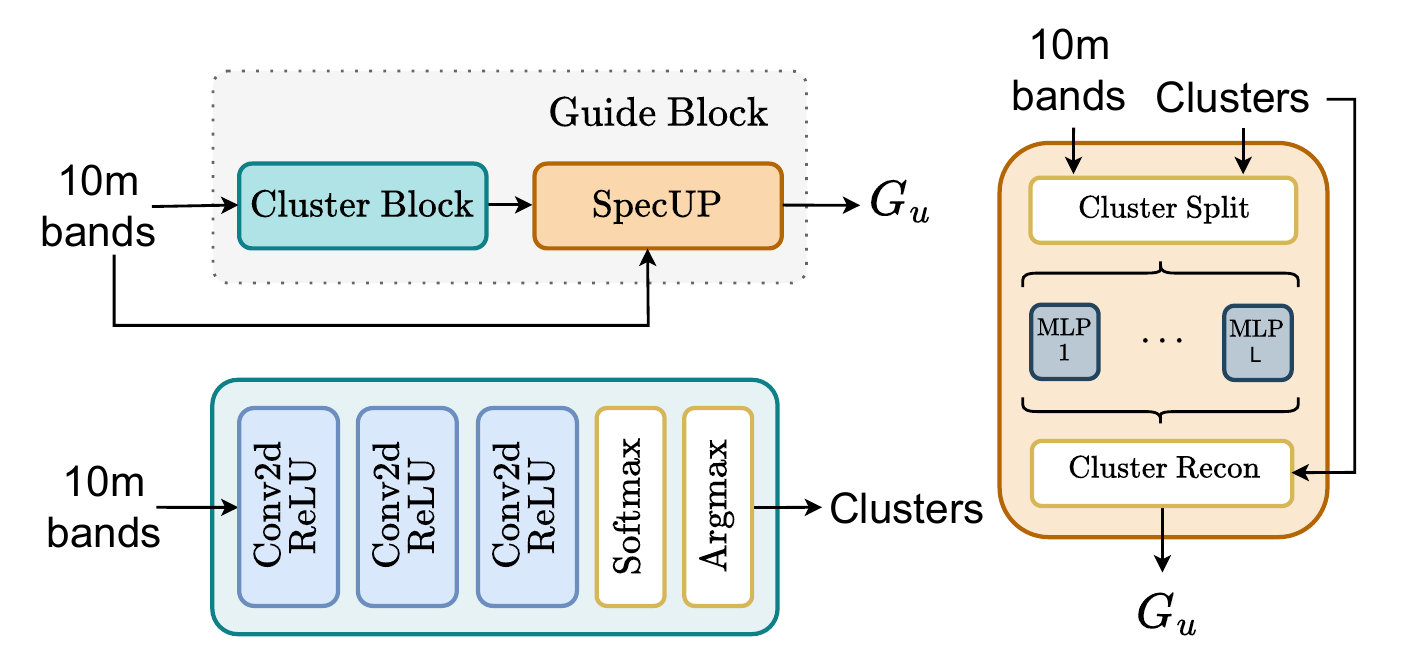}
  \caption{On the top-left, the overall module that first generates the clusters and then the guiding image. The Cluster Block is depicted in the bottom-left. On the right, the architecture of the SpecUp module that creates the guiding image $G_u$ is shown. A Multi-Layer Perceptron (MLP) is applied separately to each cluster. Cluster Split carries out the separation and flattening of the pixels according to their assigned cluster, while Cluster Recon represents the rearrangement of the pixels into the spectral domain of the 20m bands.}
  \label{fig:clusters}
\end{figure}

The second strategy employs a cluster-based learning procedure inspired by \citep{2025SSSR}, as illustrated in Figure \ref{fig:clusters}. First, we divide the domain of the 10m bands into $L$ clusters (we choose $L=5$). This clustering is performed by a sequence of 2D convolutions that map each pixel to an $L$-dimensional space, interpreted as a cluster space. The probability of each pixel belonging to a given cluster is computed using a Softmax function. Finally, an Argmax function is applied to assign each pixel to the cluster with the highest probability.

Once the clusters are computed, they are fed, together with the image composed of the 10m bands, into a module, which we call SpecUp, that constructs the guiding image. The module begins by separating and flattening the pixels according to their assigned cluster. Then, the pixels in each cluster are mapped to the spectral space of the 20m bands using an MLP with cluster-specific weights, and the resulting reconstruction is defined as the guiding image.


\subsection{Implementation details}

The proposed GINet and GINet+ models are trained over 1500 epochs using the L1 loss function between the output and the reference image. We use Adam optimizer \citep{kingma2014adam} with a learning
rate of $10^{-4}$. The number of stages is set to $K=6$, and the window size for searching similarities in the nonlocal module is set to $5\times 5$ for all attention heads. For weight computation, a $3\times $3 patch representation is used in the head-attention layers that take as input the error and the guiding image, whereas a pixel-wise representation is adopted in the layer that operates on their concatenation. In Section \ref{sec:Ablation}, we show that these are optimal choices. The guiding image module and the 6 stages of the unfolded back-projection algorithm are sequentially combined and trained jointly in an end-to-end fashion.

At the end of each epoch during training, the models are evaluated on the validation set. If the PSNR improves upon the previous best value, the parameters are updated accordingly. Thus, the final model corresponds to the one with the highest PSNR on the validation set.

Finally, we detail the number of learnable parameters in the proposed models. For each module, Guide Block, $\uparrow$, $\DB$, and $\mathcal{R}es_{\mathrm{NL}}$, the number of parameters is 29.7K, 54, 54,  and 1.0M, respectively. This results in a total of 6.3M parameters across 6 stages.

\section{Dataset and spectral descriptors}\label{sec:dataset}

In this section, we describe the Sentinel-2 dataset used for training and testing, detailing their geographic locations and data preparation procedures. Additionally, we introduce key spectral composites and indices based on 20m bands, widely employed to characterize urban, rural, and coastal environments in remote sensing applications.

\subsection{Data collection}

To derive the data used in this paper, several Sentinel-2 Level-2A products were obtained from the \href{https://www.copernicus.eu/en/access-data/conventional-data-access-hubs}{Copernicus Open Access Hub}. These products provide orthorectified, geo-referenced surface reflectance in UTM/WGS84 projection, with sub-pixel multispectral registration accuracy. Each product corresponds to a (roughly) 110km $\times$ 110km tile \citep{sentinel2-wiki}. The selected scenes were captured between January 10 and May 10, 2025, and a maximum of $1\%$ total cloud coverage was imposed.

\begin{figure}[t]
  \centering
  \begin{tabular}{cc}
    \begin{subfigure}[b]{0.55\linewidth}
      \centering
      \includegraphics[width=0.75\linewidth, trim={28cm 28cm 0 0}, clip]{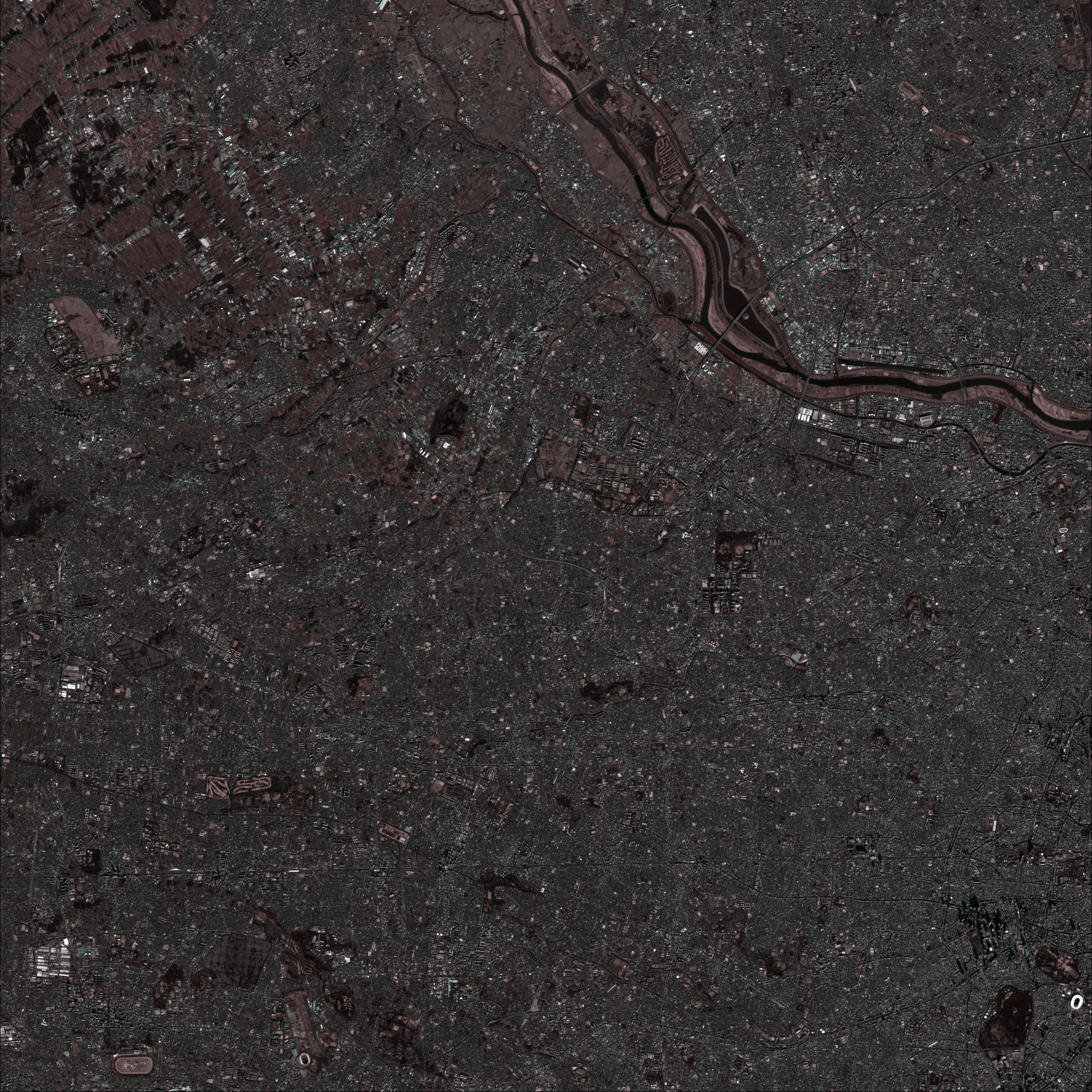}
      \caption{}
      \label{fig:subfig-a}
    \end{subfigure} & \begin{subfigure}[b]{0.275\linewidth}
      \centering
      \includegraphics[width=\linewidth, trim={14cm 14cm 0 0}, clip]{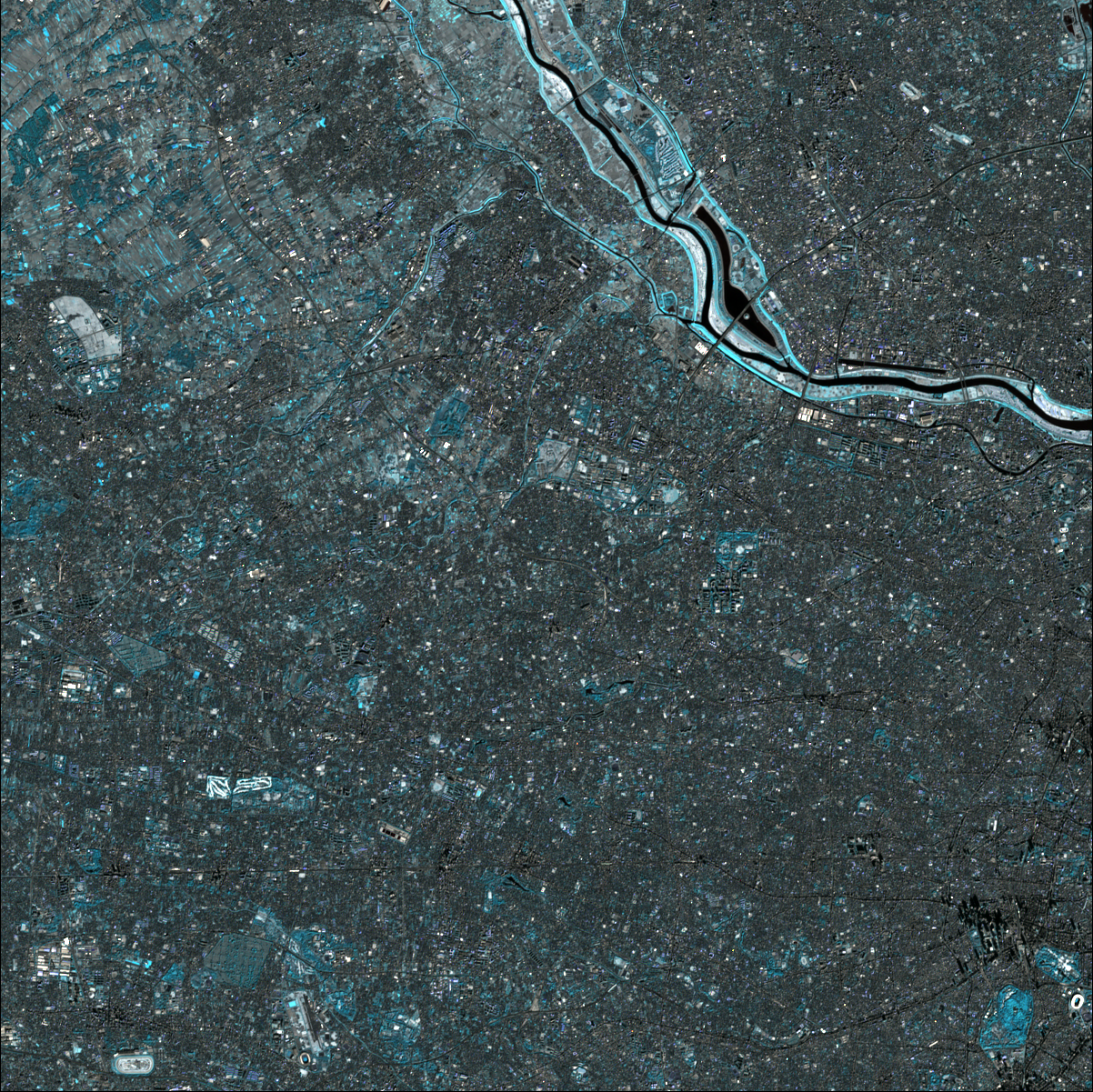}
      \caption{}
      \label{fig:subfig-b}
    \end{subfigure}
  \end{tabular}
  \caption[Area sampled from Tokyo (Japan) in the Sentinel-2 training set]{Area sampled from Tokyo (Japan) included in the Sentinel-2 training set. (a) 10m data with B4, B3 and B2 as RGB. (b) 20m data with B5, B6 and B7 as RGB.}
  \label{fig:datashowcase}
\end{figure}

The areas for training were sampled from Los Angeles (USA), New York (USA), Paris (France), Tokyo (Japan), and Guangzhou (China), while the validation area was selected from Delhi (India). Each area was further divided into 100 non-overlapping crops of size 2.4km $\times$ 2.4km. This yield, for each crop, four 10m bands of size 240 $\times$ 240 pixels, and six 20m bands of size 120 $\times$ 120 pixels. True and false color composites of the sample from Tokyo are shown in Figure \ref{fig:datashowcase}.

Nine additional 24km $\times$ 24km areas were obtained from S2 Level-2A products under the same conditions, taken over urban, rural and coastal regions. Three urban areas were sampled from Buenos Aires (Argentine), Prague (Czech Republic), and Rome (Italy); three rural areas from the Northern Territory of Australia (specifically an area encompassing the Arnold river), Whatatutu (New Zealand), and Gmina Gruta (Poland); and three coastal areas from Palma (Spain), Athens (Greece), and Barcelona (Spain). The same crop-extraction procedure was applied, yielding 300 2.4km $\times$ 2.4km crops for each type of region. The resulting 900 crops are used to test the methods over differing landscapes. We refer to these three sets as Urban, Rural, and Coastal testing sets, respectively.

To compare the effectiveness of the different models, we follow Wald's protocol \citep{hallada1983image}. In this setting, the 10m and 20m data are downsampled by a factor of two, yielding images at 20m and 40m resolutions, respectively. The task then consists of enhancing the resolution of the 40m bands to 20m, using the original 20m bands as the corresponding ground truth. Additionally, the downsampled 10m bands are used to generate the guiding images in both GINet and GINet+ proposed models.

\subsection{Composites and spectral indexes}

Let us discuss some wavelengths, composites, and indexes involving 20m S2 bands that are widely used for remote sensing in urban, rural, and coastal settings. To begin with, the \textit{urban false color composite}, which uses B12, B11, and B4 bands as RGB, highlights urbanized areas in white, gray, or purple hues, in contrast to vegetation-rich regions, which appear in green tones. Moreover, water bodies and flooded areas are also easily distinguishable in this composite, as they appear in black or dark blue \citep{sentinel2-wiki}.

In rural areas, B5, B6, B7, and B8a bands fall within the VNIR vegetation red edge spectral domain. This spectral range is widely used for the study of vegetation, as it contains the wavelengths at which chlorophyll exhibits a sharp increase in reflectance \citep{Seager2005VIR}. The \textit{SWIR composite}, which uses B12, B8a, and B4 bands as RGB, is utilized to estimate how much water is present in vegetation and soil, as water absorbs SWIR wavelengths, and also to map fire damages, since SWIR bands are very sensitive to freshly burnt land \citep{sentinel2-wiki,kato2021swir}. In a similar vein, the \textit{Normalized Difference Moisture Index} (NDMI), which is computed using B8a and B11 bands\footnote{By spectral proximity, replacing B8a by B8 yields similar values.} as 
$$
\text{NDMI}=\frac{\mbox{B8A}-\mbox{B11}}{\mbox{B8A}+\mbox{B11}},
$$
more accurately measures the water content of vegetation by mitigating the influence of internal leaf structure and dry matter content, and is employed for the monitoring of droughts \citep{sentinel2-wiki,GAO1996NDWI}.

For coastal regions, we highlight the \textit{Normalized Difference Water Index} (NDWI), which involves B3 and B8a bands\footnote{Again, the same expression with B8 instead of B8a gives a comparable index.} as follows:
\begin{equation}\label{s2-ndwi}
\text{NDWI}=\frac{\mbox{B3}-\mbox{B8a}}{\mbox{B3}+\mbox{B8a}}.
\end{equation}
Nowadays, this index is mainly used to detect and monitor slight changes in the water content of water bodies \citep{sentinel2-wiki,McFEETERS01051996}.

\begin{sidewaystable*}
\centering
\caption{Quantitative metrics obtained by each method on the Sentinel-2 validation and testing sets. The best values are highlighted in bold, the second-best are underlined, and the third-best are in italic. The proposed GINet+ and GINet achieve the best and second-best results on both sets and across all metrics. WINet ranks third on validation (except for SAM), while MMNet holds this position on testing across all metrics.}
\begin{tabular}{l |c c |c c |c c |c c}
 & \multicolumn{2}{c}{ERGAS$\downarrow$}& \multicolumn{2}{c}{PSNR$\uparrow$} & \multicolumn{2}{c}{SSIM$\uparrow$} & \multicolumn{2}{c}{SAM$\downarrow$} \\
 & validation & test & validation & test & validation &test  & validation & test \\
    \hline                          
    Bicubic                         & 0.9830 & 1.4621 & 34.20  & 35.26 & 0.8694 & 0.8899 & 1.0706& 1.2772\\
 PCA& 4.5922& 4.1492& 24.93& 26.19& 0.8720& 0.8297& 9.6154&8.5721\\
 GSA& 1.7866& 1.2693& 33.51& 36.05& 0.9098& 0.9130& 2.4855&1.3838\\
 CNMF& 2.1183& 1.4199& 31.26& 34.83& 0.8918& 0.9002& 2.6183&1.4661\\  \hline
    DSen2                           & 0.4109 & 0.6395 & 41.81  & 41.66 & 0.9757 & 0.9754 & \textit{0.6474} & 0.8696\\
    ResSen2                         & 0.4477 & 0.6892 & 41.06  & 40.80 & 0.9718 & 0.9711 & 0.6990& 0.8938\\
    FusionNet& 0.4910& 0.8760& 40.27& 38.60& 0.9669& 0.9581& 0.7100&0.9887\\  
    SRPPNN                          & 0.4130 & 0.6484 & 41.77  & 41.12 & 0.9762 & 0.9741 & 0.6646& 0.8855\\
    MMNet                           & 0.4037 & \textit{0.6022} & 41.96  & \textit{41.90} & 0.9771 & \textit{0.9773} & 0.6646& \textit{0.8461}\\
    AWFLN                           & 0.4522 & 0.7251 & 40.95  & 40.02 & 0.9737 & 0.9699 & 0.6990& 0.9486\\
    UTeRM                           & 0.8597 & 1.2849 & 35.68  & 34.17 & 0.9429 & 0.9492 & 1.4782& 1.8933\\
    LGCT                            & 0.4126 & 0.6356 & 41.78  & 41.31 & 0.9765 & 0.9755 & 0.6704& 0.8964\\
    WINet                           & \textit{0.3990} & 0.6608 & \textit{42.05}  & 40.57 & \textit{0.9776} & 0.9730 & 0.6532& 0.8779\\ \hline
    GINet  & \underline{0.3757} & \underline{0.5747} & \underline{42.59}  & \underline{42.52} & \underline{0.9798} & \underline{0.9791} & \underline{0.6188} & \underline{0.8155}\\
    GINet+     & \textbf{0.3643} & \textbf{0.5546} & \textbf{42.84}  & \textbf{42.85} & \textbf{0.9806} & \textbf{0.9809} & \textbf{0.6016} & \textbf{0.7932} \\
\end{tabular}
\label{tab:QuantVal}
\end{sidewaystable*}

\section{Experimental results} \label{sec:ExpResults}

In this section, we evaluate the performance of the proposed GINet and GINet+ models for enhancing the resolution of 20m Sentinel-2 bands. We use the dataset described in the previous section, which consists of 500 crops for training and 100 for validation. The testing set comprises 900 crops, uniformly distributed across Urban, Rural, and Coastal subsets.

For the quantitative comparison, we use the following metrics: PSNR (Peak Signal to Noise Ratio) \citep{rabbani1991digital}, which measures the spatial reconstruction quality with respect to noise; SSIM (Structural Similarity Index Measure) \citep{wang2002universal}, which assesses the general quality of the fused image through the loss of correlation, luminance, and contrast distortion; SAM (Spectral Angle Mapper) \citep{alparone2004global}, which evaluates the spectral reconstruction quality; and ERGAS (Error Relative Globale Adimentionelle) \citep{rabbani1991digital,ranchin2000fusion},  which estimates the global spatial quality.

We use bicubic interpolation as a baseline and compare with the classic fusion methods GSA \citep{dai2007bilateral} and CNMF \citep{yokoya2011coupled}; the S2-specific learning-based approaches DSen2 \citep{2018-lanaras-network} and ResSen2 \citep{ResNet2018}; and the learning-based fusion methods FusionNet \citep{deng2020detail}, SRPPNN \citep{cai2020super}, MMNet \citep{yan2022mmnet}, AWFLN \citep{lu2023awfln}, UTeRM \citep{mai2024deep}, LGCT \citep{2024LGCT}, and WINet \citep{2024WINet}. The implementation codes for the classic approaches were obtained from the \href{https://github.com/codegaj/py_pansharpening/tree/master}{Py\_pansharpening toolbox}, while FusionNet was sourced from the \href{https://github.com/liangjiandeng/DLPan-Toolbox/tree/main}{DLPan-toolbox}. DSen2 and ResSen2 were implemented by us, and the remaining methods were downloaded from the websites of the authors.

\begin{table*}\centering\caption{Quantitative metrics obtained by each method on the Urban Sentinel-2 testing set. The best values are highlighted in bold, the second-best are underlined, and the third-best are in italic. The proposed GINet+ and GINet achieve the best and second-best results across all metrics, while MMNet ranks third.}
\begin{tabular}{l |c | c | c | c}
    & ERGAS$\downarrow$& PSNR$\uparrow$ & SSIM$\uparrow$ & SAM$\downarrow$ \\
    \hline                          
    Bicubic                         & 2.1496 & 32.05 & 0.8387 & 1.9720\\
 PCA& 4.8888& 25.17& 0.8086&10.3653\\
 GSA& 1.2574& 35.92& 0.9259&1.4898\\
 CNMF& 1.4650& 33.89& 0.9052&1.6294\\ \hline
    DSen2                           & 0.9211 & 39.83 & 0.9695 & 1.3369\\
    ResSen2                         & 0.9754 & 39.29 & 0.9661 & 1.3484\\
 FusionNet& 1.2186& 37.67& 0.9470&1.4610\\ 
    SRPPNN                          & 0.8959 & 40.04 & 0.9716 & 1.3159\\
    MMNet                           & \textit{0.8433} & \textit{40.57} & \textit{0.9745} & \textit{1.2681} \\
    AWFLN                           & 0.9901 & 39.21 & 0.9665 & 1.3655\\
    UTeRM                           & 1.4561 & 35.85 & 0.9426 & 2.0760\\
    LGCT                            & 0.8654 & 40.39 & 0.9736 & 1.2987\\
    WINet                           & 0.8515 & 40.48 & 0.9742 & 1.2720\\ \hline
    GINet  & \underline{0.8111} & \underline{40.93} & \underline{0.9763} & \underline{1.2280}\\
    GINet+     & \textbf{0.7949} & \textbf{41.11} & \textbf{0.9773} & \textbf{1.2070}\\
\end{tabular}
\label{tab:QuantUrban}
\end{table*}

For all fusion methods, the PAN images were derived using the spectral-similarity strategy presented in \citep{gavsparovic2018effect}, as done in our GINet model. The state-of-the-art techniques based on deep learning were trained using the loss functions and hyperparameters specified in their respective articles. The weights used for the evaluation correspond to the learnable parameters achieving the best performance in terms of PSNR on the validation set.

\subsection{Quantitative results}

In Table~\ref{tab:QuantVal}, we report the quantitative metrics obtained by each method on both the validation and full testing sets. The proposed GINet+ and GINet models achieve the best and second-best performance, respectively, across all metrics in both sets. WINet ranks third in the validation set but shows limited generalization capability, as its performance drops considerably in testing. In contrast, MMNet displays more consistent behavior, securing third place in the testing set. DSen2 clearly outperforms ResSen2 and yields competitive results, ranking just behind our models and MMNet.

\begin{table*}\centering\caption{Quantitative metrics obtained by each method on the Rural Sentinel-2 testing set. The best values are highlighted in bold, the second-best are underlined, and the third-best are in italic. The proposed GINet+ and GINet achieve the best and second-best results across all metrics, while MMNet ranks third.}
\begin{tabular}{l |c | c | c | c}
    & ERGAS$\downarrow$& PSNR$\uparrow$ & SSIM$\uparrow$ & SAM$\downarrow$ \\
    \hline                          
    Bicubic                         & 1.1699 & 34.15 & 0.9040 & 1.0891\\
 PCA& 5.0125& 24.21& 0.8025&10.1814\\
 GSA& 1.6513& 34.13& 0.8877&1.7977\\
 CNMF& 1.7780& 33.16& 0.8731&1.8959\\ \hline
    DSen2                           & 0.4699 & 41.75 & 0.9814 & 0.6895\\
    ResSen2                         & 0.5121 & 40.92 & 0.9769 & 0.7238\\
 FusionNet& 0.6580& 38.93& 0.9678&0.8317\\  
    SRPPNN                          & 0.4822 & 41.69 & 0.9813 & 0.7200\\
    MMNet                           & \textit{0.4448} & \textit{42.14} & \textit{0.9826} & \textit{0.6818}\\
    AWFLN                           & 0.5544 & 40.27 & 0.9768 & 0.7907\\
    UTeRM                           & 1.0529 & 34.87 & 0.9586 & 1.7322\\
    LGCT                            & 0.4815 & 41.50 & 0.9813 & 0.7353\\
    WINet                           & 0.5171 & 40.81 & 0.9785 & 0.7448\\ \hline
    GINet  & \underline{0.4202} & \underline{42.58} & \underline{0.9841} & \underline{0.6570}\\
    GINet+     & \textbf{0.3962} & \textbf{43.16} & \textbf{0.9863} & \textbf{0.6283}\\
\end{tabular}
\label{tab:QuantRural}
\end{table*}

Tables~\ref{tab:QuantUrban}, \ref{tab:QuantRural}, and \ref{tab:QuantCoastal} detail the results for the Urban, Rural, and Coastal subsets of the testing data, respectively. Once again, GINet+ and GINet outperform all other methods, securing the top two positions across all metrics in each landscape. Among the remaining approaches, MMNet and DSen2 deliver the strongest results. MMNet ranks third across all metrics in the Urban and Rural subsets, while DSen2 takes third place in the Coastal subset, outperforming MMNet in all metrics except ERGAS.

The relative performance of competing methods varies across different landscapes. MMNet tends to be more robust in urban scenes, likely due to its unfolding-based architecture, which better captures sharp structures and geometric variations. In contrast, DSen2 shows improved results in more homogeneous areas such as coastal regions, where its design favors smoother reconstructions and spectral consistency. Our GINet and GINet+ models consistently outperform the others across all environments, suggesting that their hybrid attention mechanisms and multi-stage structure are well suited to the diverse spatial and spectral characteristics of S2 imagery. This underlines the importance of designing architectures specifically adapted to the properties of the sensor and the variability of real-world scenes.

\begin{table*}\centering\caption{Quantitative metrics obtained by each method on the Coastal Sentinel-2 testing set. The best values are highlighted in bold, the second-best are underlined, and the third-best are in italic. The proposed GINet+ and GINet achieve the best and second-best results across all metrics, while DSen2 ranks third (except for ERGAS).}
\begin{tabular}{l |c | c | c | c}
    & ERGAS$\downarrow$& PSNR$\uparrow$ & SSIM$\uparrow$ & SAM$\downarrow$ \\
    \hline                          
    Bicubic                         & 1.0669 & 39.59 & 0.9271 & 0.7706\\
 PCA& 2.5464& 29.20& 0.8780&5.1698\\
 GSA& 0.8993& 38.11& 0.9253&0.8637\\
 CNMF& 1.0168& 37.43& 0.9223&0.8728\\ \hline
    DSen2                           & 0.5275 & \textit{43.39} & \textit{0.9754} & \textit{0.5825}\\
    ResSen2                         & 0.5803 & 42.19 & 0.9704 & 0.6092\\
 FusionNet& 0.7515& 39.21& 0.9594&0.6736\\ 
    SRPPNN                          & 0.5670 & 41.63 & 0.9695 & 0.6207\\
    MMNet                           & \textit{0.5184} & 42.98 & 0.9747 & 0.5882\\
    AWFLN                           & 0.6307 & 40.58 & 0.9663 & 0.6895\\
    UTeRM                           & 1.3457 & 31.79 & 0.9465 & 1.8717\\
    LGCT                            & 0.5599 & 42.05 & 0.9716 & 0.6551\\
    WINet                           & 0.6138 & 40.97 & 0.9662 & 0.6169\\
 \hline
    GINet  & \underline{0.4927} & \underline{44.04} & \underline{0.9768} & \underline{0.5615}\\
    GINet+     & \textbf{0.4728} & \textbf{44.28} & \textbf{0.9792} & \textbf{0.5443}\\
\end{tabular}
\label{tab:QuantCoastal}
\end{table*}

\subsection{Qualitative results}

\begin{figure}[t]
    \centering
    \begin{minipage}[b]{0.31\columnwidth}
        \includegraphics[width=\linewidth, clip, trim={5cm 9cm 17cm 13cm}]{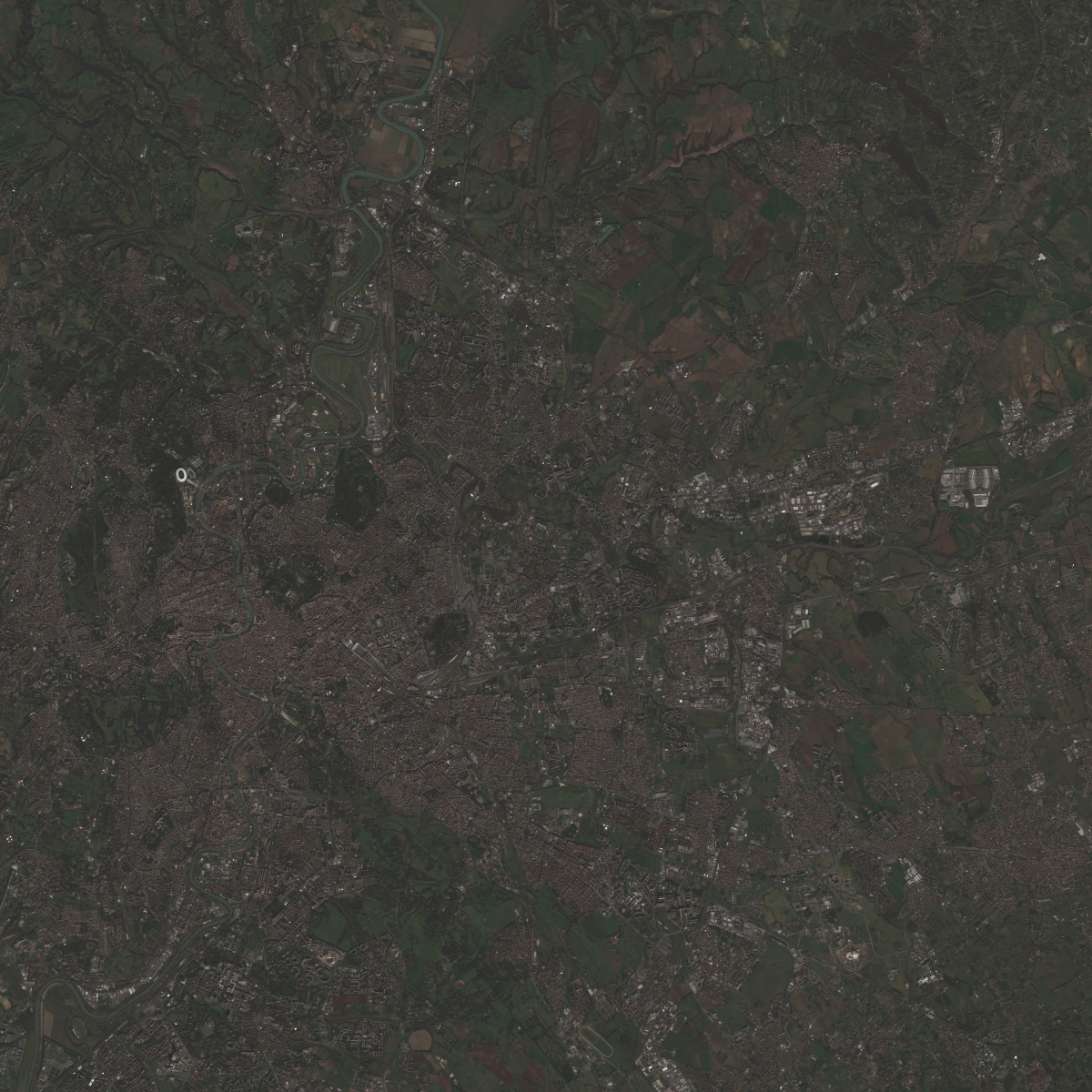}
        \vspace{-6mm}
        \caption*{Rome}
    \end{minipage}
    \hspace{-1mm}
    \begin{minipage}[b]{0.31\columnwidth}
        \includegraphics[width=\linewidth, clip, trim={2cm 2cm 6cm 6cm}]{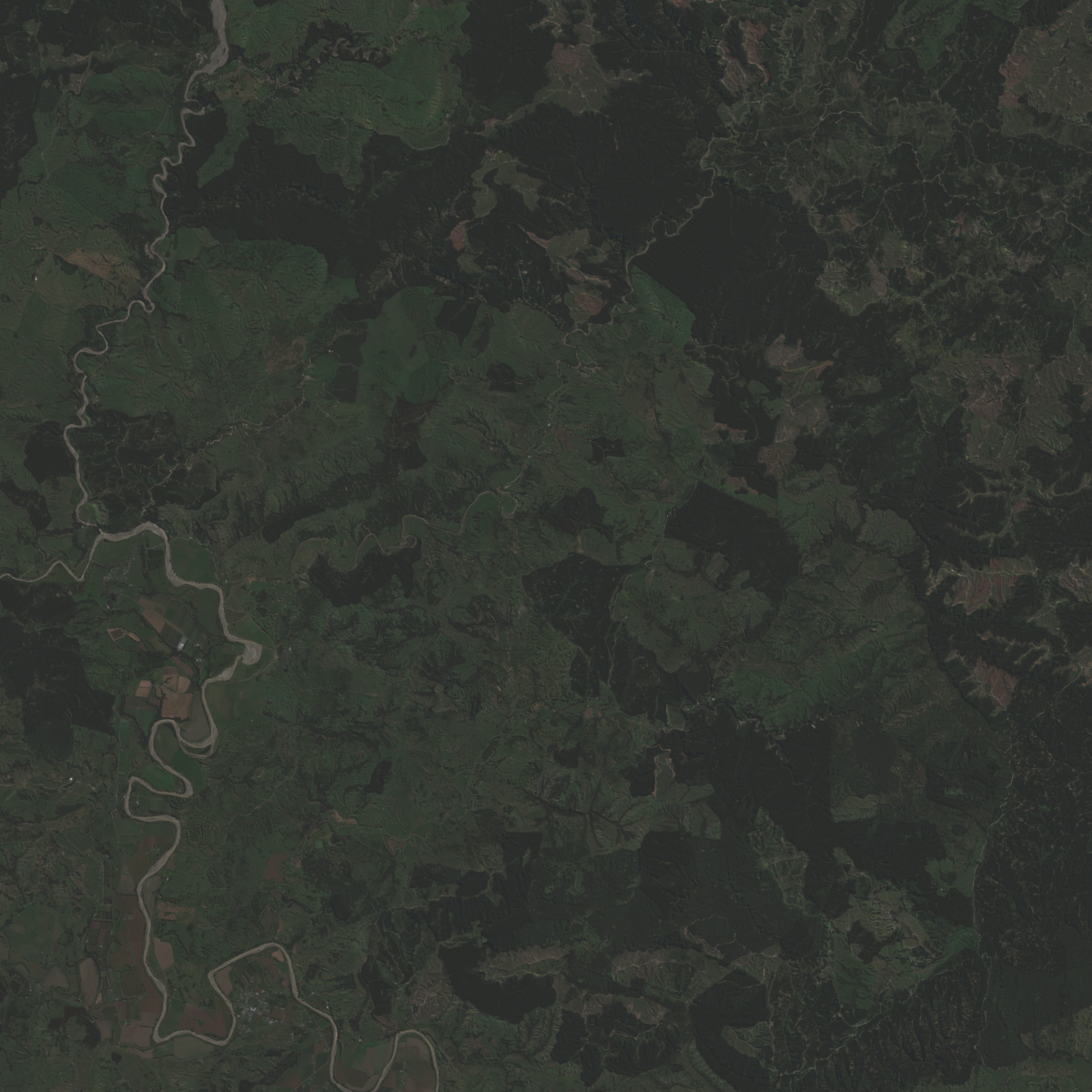}
        \vspace{-6mm}
        \caption*{New Zealand}
    \end{minipage}
    \hspace{-1mm}
    \begin{minipage}[b]{0.31\columnwidth}
        \includegraphics[width=\linewidth, clip, trim={4cm 4cm 7cm 7cm}]{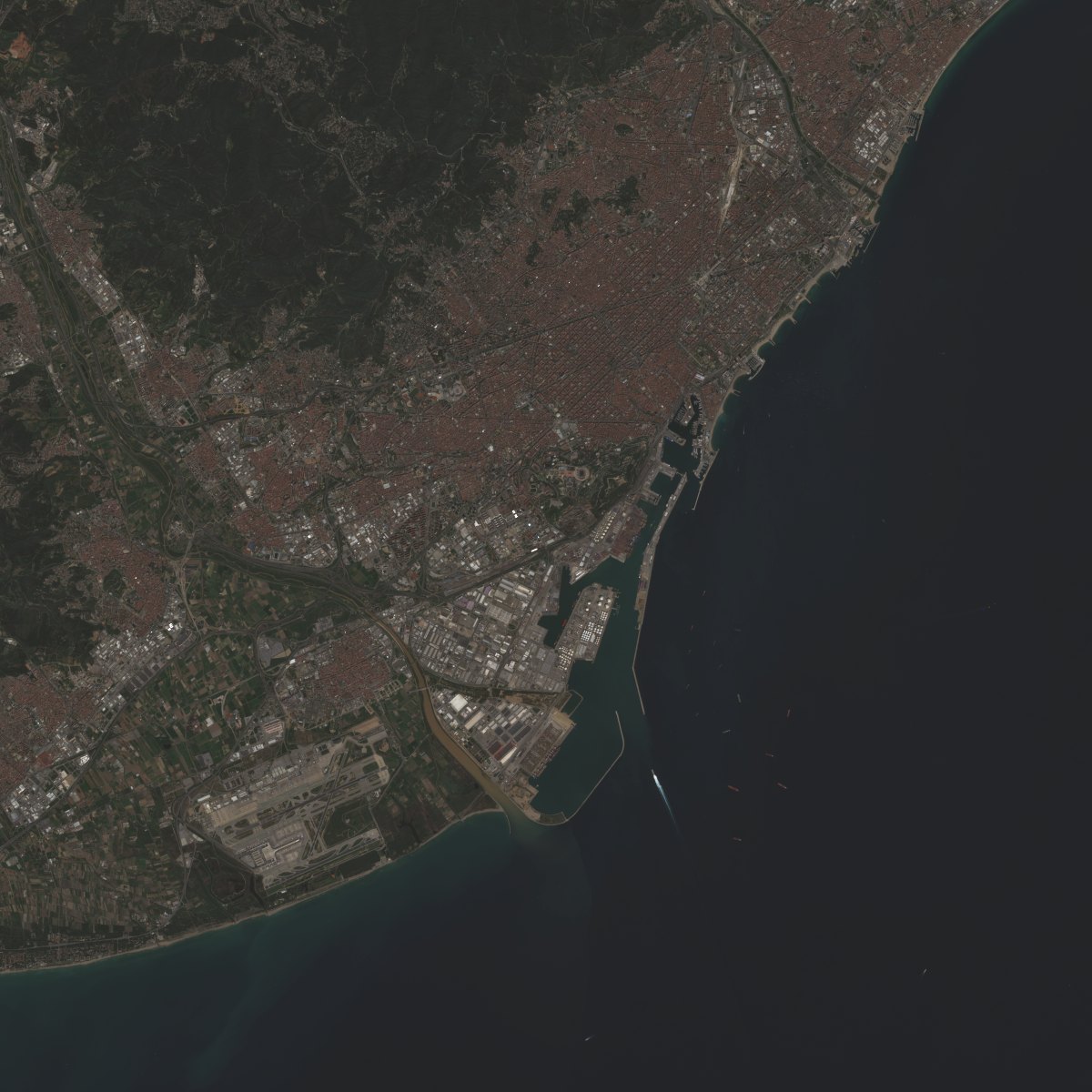}
        \vspace{-6mm}
        \caption*{Barcelona}
    \end{minipage}
    \caption[True color composites of the Sentinel-2 test images selected for visual comparison]{True color composites (using B4, B3, and B2 bands as RGB) of the Sentinel-2 test images selected for visual comparison.}
    \label{imgs:visual-comparison}
\end{figure}

To assess visual performance, we use an urban, a rural and a coastal crop, extracted from Rome, New Zealand, and Barcelona, respectively. Their true color composites are shown in Figure~\ref{imgs:visual-comparison}. For each enhanced image, white balance and gamma correction are applied to each channel for better visualization. Additionally, we present the error maps computed as the mean absolute difference between the output provided by each method and the ground truth across all channels. These maps have been clipped and rescaled to highlight the errors. Due to space limitations and the poor performance of PCA across all metrics, its results have been omitted in this subsection.

\begin{figure}[p]
    \centering
    \begin{subfigure}[b]{0.18\textwidth}
        \centering
        \spyKorea{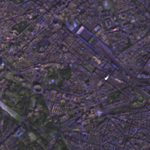}
        \spyKorea{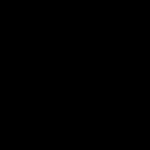}
        \captionsetup{font=small, labelformat=empty}
        \caption{Reference}
    \end{subfigure}
    \begin{subfigure}[b]{0.18\textwidth}
        \centering
        \spyKorea{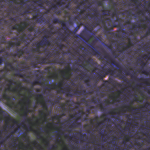}
        \spyKorea{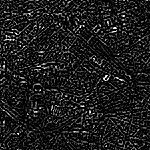}
        \captionsetup{font=small, labelformat=empty}
        \caption{Bicubic}
    \end{subfigure}
    \begin{subfigure}[b]{0.18\textwidth}
        \centering
        \spyKorea{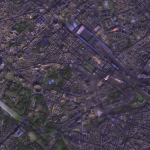}
        \spyKorea{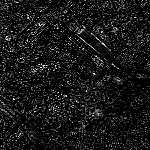}
        \captionsetup{font=small, labelformat=empty}
        \caption{CNMF}
    \end{subfigure}
    \begin{subfigure}[b]{0.18\textwidth}
        \centering
        \spyKorea{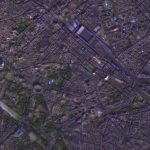}
        \spyKorea{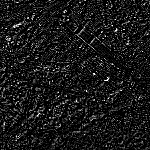}
        \captionsetup{font=small, labelformat=empty}
        \caption{GSA}
    \end{subfigure}
    \begin{subfigure}[b]{0.18\textwidth}
        \centering
        \spyKorea{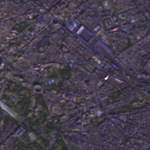}
        \spyKorea{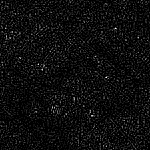}
        \captionsetup{font=small, labelformat=empty}
        \caption{DSen2}
    \end{subfigure}

    \begin{subfigure}[b]{0.18\textwidth}
        \centering
        \spyKorea{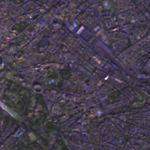}
        \spyKorea{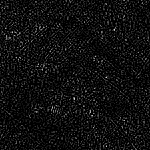}
        \captionsetup{font=small, labelformat=empty}
        \caption{ResSen2}
    \end{subfigure}
    \begin{subfigure}[b]{0.18\textwidth}
        \centering
        \spyKorea{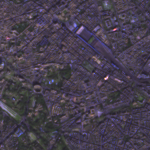}
        \spyKorea{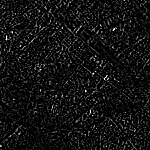}
        \captionsetup{font=small, labelformat=empty}
        \caption{FusionNet}
    \end{subfigure}
    \begin{subfigure}[b]{0.18\textwidth}
        \centering
        \spyKorea{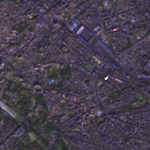}
        \spyKorea{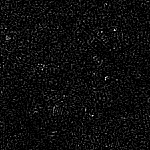}
        \captionsetup{font=small, labelformat=empty}
        \caption{SRPPNN}
    \end{subfigure}
    \begin{subfigure}[b]{0.18\textwidth}
        \centering
        \spyKorea{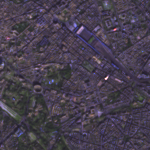}
        \spyKorea{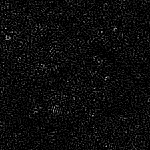}
        \captionsetup{font=small, labelformat=empty}
        \caption{MMNet}
    \end{subfigure}
    \begin{subfigure}[b]{0.18\textwidth}
        \centering
        \spyKorea{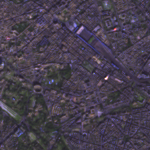}
        \spyKorea{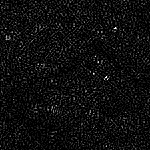}
        \captionsetup{font=small, labelformat=empty}
        \caption{AWFLN}
    \end{subfigure}

    \begin{subfigure}[b]{0.18\textwidth}
        \centering
        \spyKorea{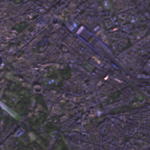}
        \spyKorea{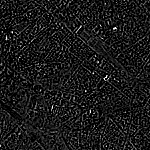}
        \captionsetup{font=small, labelformat=empty}
        \caption{UTeRM}
    \end{subfigure}
    \begin{subfigure}[b]{0.18\textwidth}
        \centering
        \spyKorea{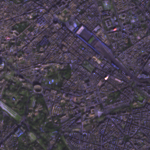}
        \spyKorea{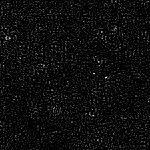}
        \captionsetup{font=small, labelformat=empty}
        \caption{LGCT}
    \end{subfigure}
    \begin{subfigure}[b]{0.18\textwidth}
        \centering
        \spyKorea{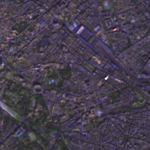}
        \spyKorea{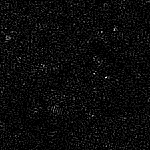}
        \captionsetup{font=small, labelformat=empty}
        \caption{WINet}
    \end{subfigure}
    \begin{subfigure}[b]{0.18\textwidth}
        \centering
        \spyKorea{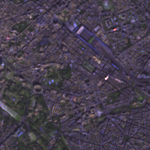}
        \spyKorea{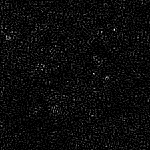}
        \captionsetup{font=small, labelformat=empty}
        \caption{GINet}
    \end{subfigure}
    \begin{subfigure}[b]{0.18\textwidth}
        \centering
        \spyKorea{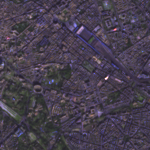}
        \spyKorea{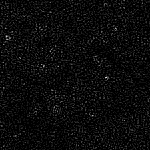}
        \captionsetup{font=small, labelformat=empty}
        \caption{GINet+}
    \end{subfigure}

    \caption[Visual results on the Urban Sentinel-2 testing set]{Visual comparison on a urban crop from Rome. Urban false color composites and error maps are shown. While some methods provide good spatial and spectral accuracy, GINet+ achieves superior color reconstruction, particularly around the highlighted red point, as further supported by the error map.
    }
    \label{fig:roma}
\end{figure}

Figure \ref{fig:roma} presents the results obtained by each method on the urban crop from Rome. Specifically, we show the urban false color composites alongside the error maps. The comparison reveals that classical techniques and FusionNet have difficulty capturing the geometry, whereas UTeRM falls short in spectral fidelity. Although the other methods yield reasonably good spatial and spectral reconstructions, our approach, GINet+, achieves more accurate color recovery, particularly evident at the red point within the highlighted area. These observations are further supported by comparing the error maps.

\begin{figure}[p]
    \centering
    \begin{subfigure}[b]{0.18\textwidth}
        \centering
        \spyPoland{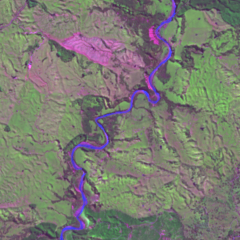}
        \spyPoland{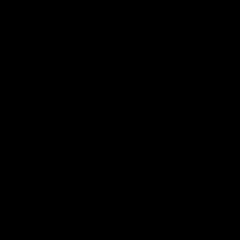}
        \captionsetup{font=small, labelformat=empty}
        \caption{Reference}
    \end{subfigure}
    \begin{subfigure}[b]{0.18\textwidth}
        \centering
        \spyPoland{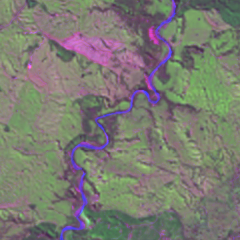}
        \spyPoland{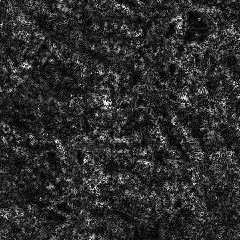}
        \captionsetup{font=small, labelformat=empty}
        \caption{Bicubic}
    \end{subfigure}
    \begin{subfigure}[b]{0.18\textwidth}
        \centering
        \spyPoland{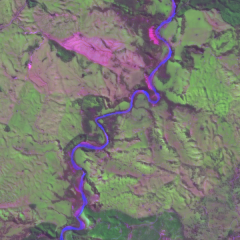}
        \spyPoland{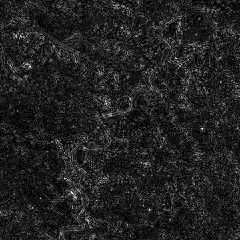}
        \captionsetup{font=small, labelformat=empty}
        \caption{CNMF}
    \end{subfigure}
    \begin{subfigure}[b]{0.18\textwidth}
        \centering
        \spyPoland{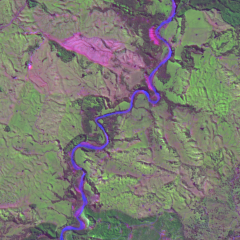}
        \spyPoland{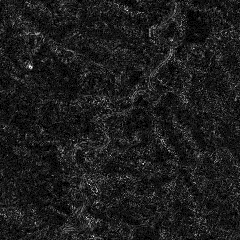}
        \captionsetup{font=small, labelformat=empty}
        \caption{GSA}
    \end{subfigure}
    \begin{subfigure}[b]{0.18\textwidth}
        \centering
        \spyPoland{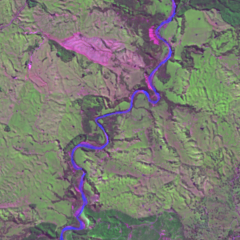}
        \spyPoland{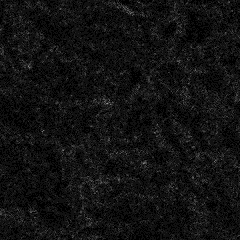}
        \captionsetup{font=small, labelformat=empty}
        \caption{DSen2}
    \end{subfigure}

    \begin{subfigure}[b]{0.18\textwidth}
        \centering
        \spyPoland{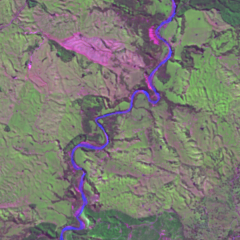}
        \spyPoland{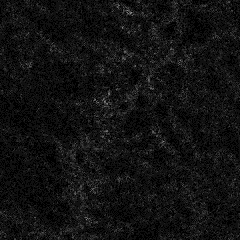}
        \captionsetup{font=small, labelformat=empty}
        \caption{ResSen2}
    \end{subfigure}
    \begin{subfigure}[b]{0.18\textwidth}
        \centering
        \spyPoland{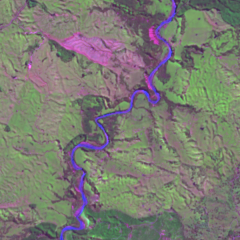}
        \spyPoland{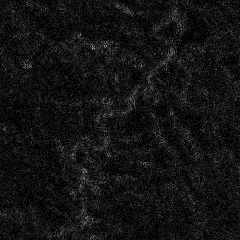}
        \captionsetup{font=small, labelformat=empty}
        \caption{FusionNet}
    \end{subfigure}
    \begin{subfigure}[b]{0.18\textwidth}
        \centering
        \spyPoland{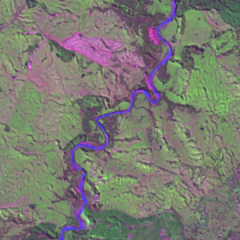}
        \spyPoland{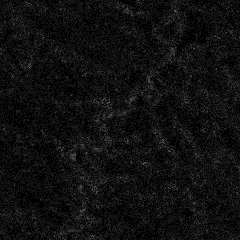}
        \captionsetup{font=small, labelformat=empty}
        \caption{SRPPNN}
    \end{subfigure}
    \begin{subfigure}[b]{0.18\textwidth}
        \centering
        \spyPoland{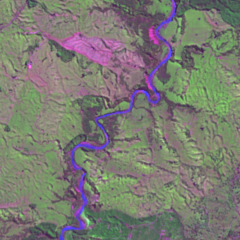}
        \spyPoland{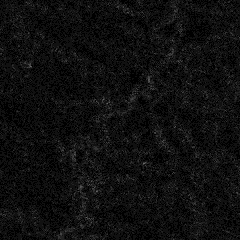}
        \captionsetup{font=small, labelformat=empty}
        \caption{MMNet}
    \end{subfigure}
    \begin{subfigure}[b]{0.18\textwidth}
        \centering
        \spyPoland{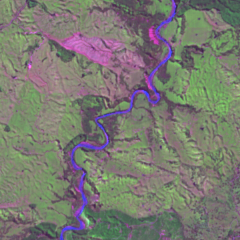}
        \spyPoland{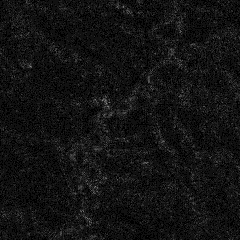}
        \captionsetup{font=small, labelformat=empty}
        \caption{AWFLN}
    \end{subfigure}

    \begin{subfigure}[b]{0.18\textwidth}
        \centering
        \spyPoland{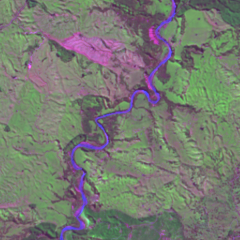}
        \spyPoland{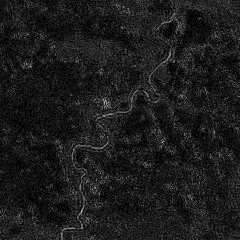}
        \captionsetup{font=small, labelformat=empty}
        \caption{UTeRM}
    \end{subfigure}
    \begin{subfigure}[b]{0.18\textwidth}
        \centering
        \spyPoland{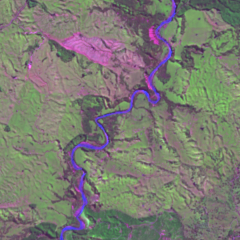}
        \spyPoland{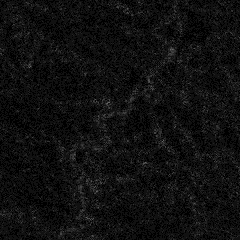}
        \captionsetup{font=small, labelformat=empty}
        \caption{LGCT}
    \end{subfigure}
    \begin{subfigure}[b]{0.18\textwidth}
        \centering
        \spyPoland{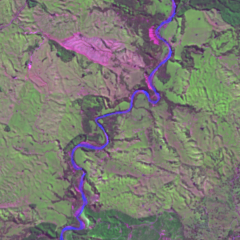}
        \spyPoland{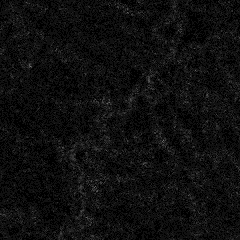}
        \captionsetup{font=small, labelformat=empty}
        \caption{WINet}
    \end{subfigure}
    \begin{subfigure}[b]{0.18\textwidth}
        \centering
        \spyPoland{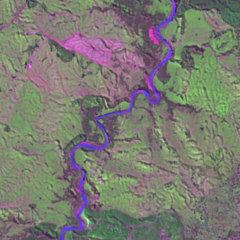}
        \spyPoland{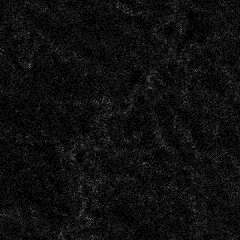}
        \captionsetup{font=small, labelformat=empty}
        \caption{GINet}
    \end{subfigure}
    \begin{subfigure}[b]{0.18\textwidth}
        \centering
        \spyPoland{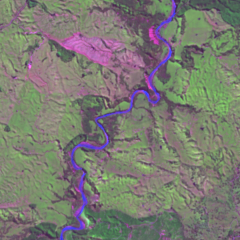}
        \spyPoland{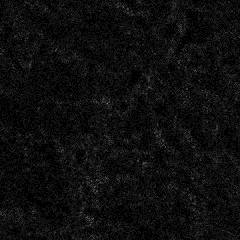}
        \captionsetup{font=small, labelformat=empty}
        \caption{GINet+}
    \end{subfigure}
    \caption[Visual results on the Rural Sentinel-2 testing set]{Visual comparison on a rural crop from New Zealand. SWIR composites and error maps are shown. Differences are difficult to discern due to the use of the 10m B4 band. However, the error maps reveal that SRPPNN, MMNet, LGCT, WINet and ours achieve more accurate reconstructions than the other methods.
    }
    \label{fig:NovaZelanda}
\end{figure}

\begin{figure}[p]
    \centering
    \begin{subfigure}[b]{0.18\textwidth}
        \centering
        \spyBCN{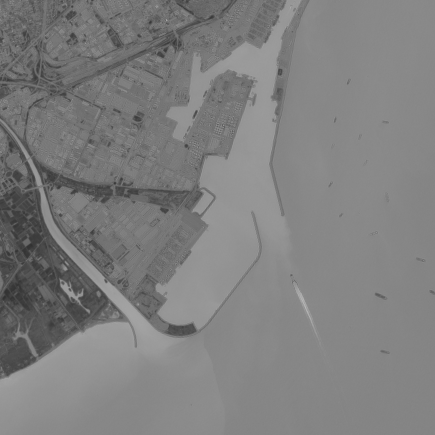}
        \spyBCN{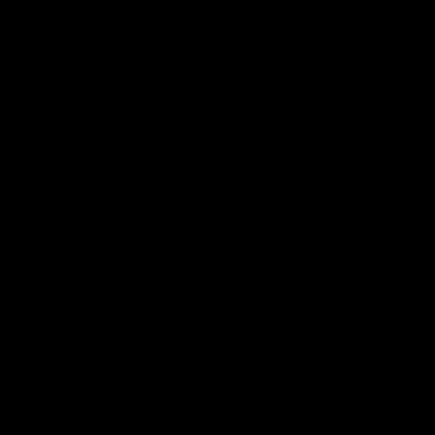}
        \captionsetup{font=small, labelformat=empty}
        \caption{Reference}
    \end{subfigure}
    \begin{subfigure}[b]{0.18\textwidth}
        \centering
        \spyBCN{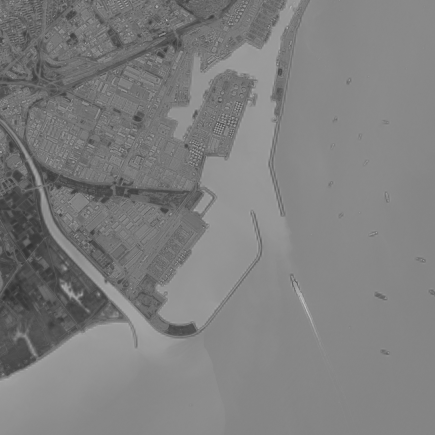}
        \spyBCN{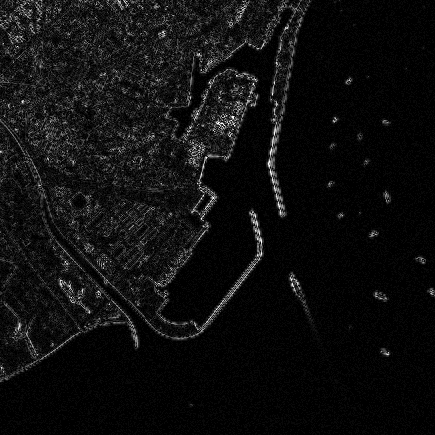}
        \captionsetup{font=small, labelformat=empty}
        \caption{Bicubic}
    \end{subfigure}
    \begin{subfigure}[b]{0.18\textwidth}
        \centering
        \spyBCN{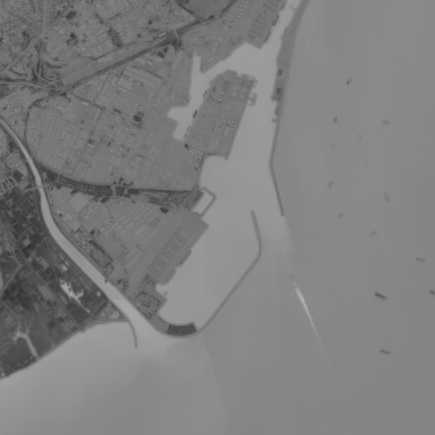}
        \spyBCN{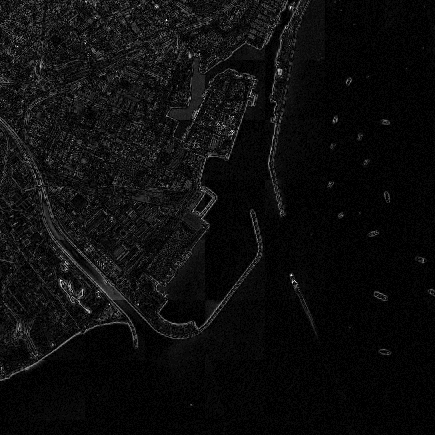}
        \captionsetup{font=small, labelformat=empty}
        \caption{CNMF}
    \end{subfigure}
    \begin{subfigure}[b]{0.18\textwidth}
        \centering
        \spyBCN{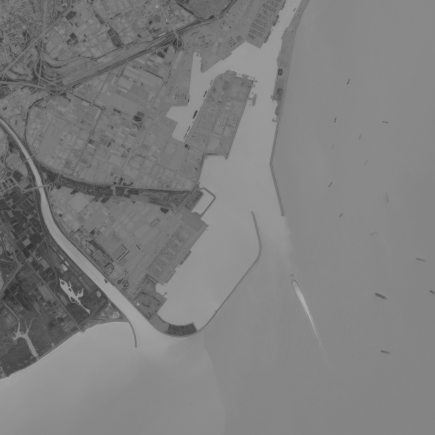}
        \spyBCN{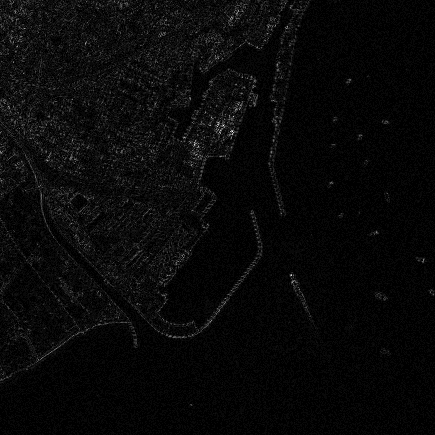}
        \captionsetup{font=small, labelformat=empty}
        \caption{GSA}
    \end{subfigure}
    \begin{subfigure}[b]{0.18\textwidth}
        \centering
        \spyBCN{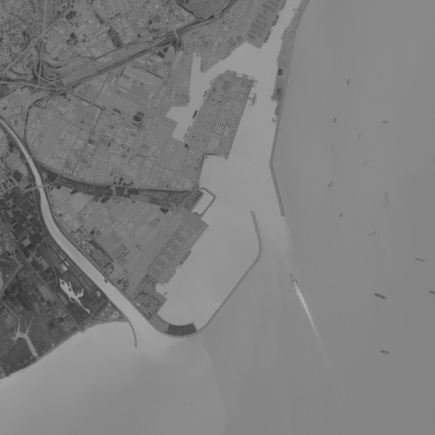}
        \spyBCN{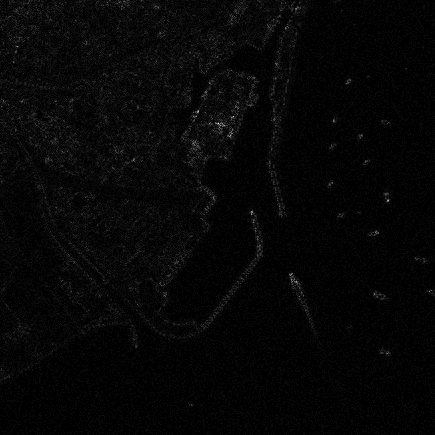}
        \captionsetup{font=small, labelformat=empty}
        \caption{DSen2}
    \end{subfigure}

    \begin{subfigure}[b]{0.18\textwidth}
        \centering
        \spyBCN{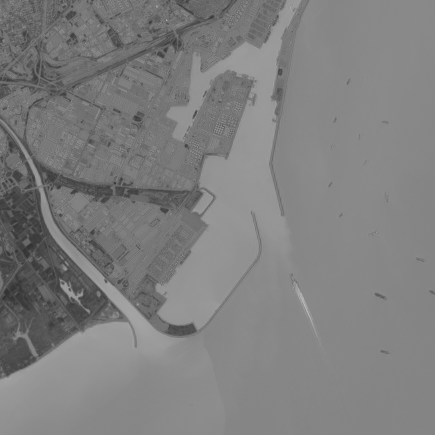}
        \spyBCN{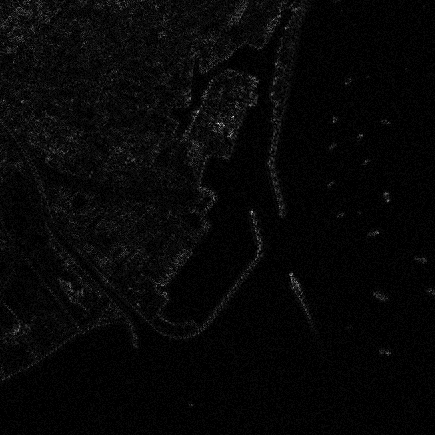}
        \captionsetup{font=small, labelformat=empty}
        \caption{ResSen2}
    \end{subfigure}
    \begin{subfigure}[b]{0.18\textwidth}
        \centering
        \spyBCN{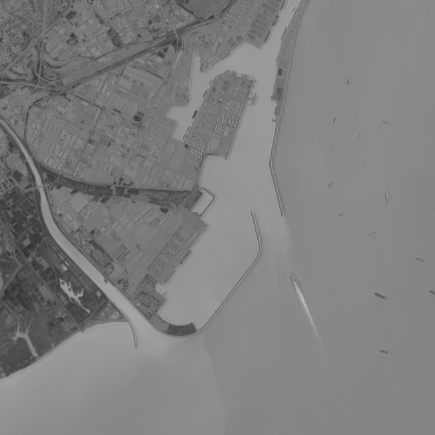}
        \spyBCN{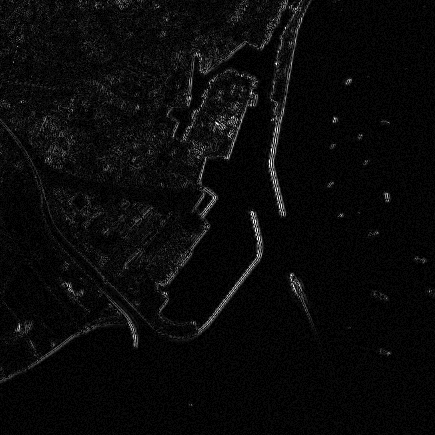}
        \captionsetup{font=small, labelformat=empty}
        \caption{FusionNet}
    \end{subfigure}
    \begin{subfigure}[b]{0.18\textwidth}
        \centering
        \spyBCN{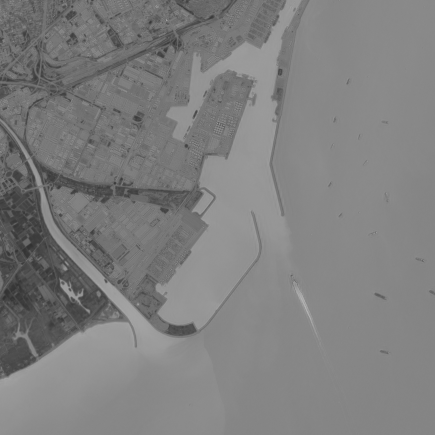}
        \spyBCN{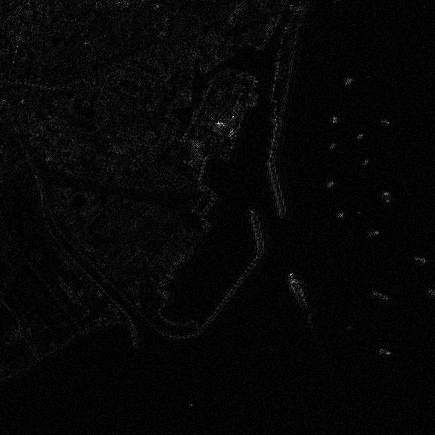}
        \captionsetup{font=small, labelformat=empty}
        \caption{SRPPNN}
    \end{subfigure}
    \begin{subfigure}[b]{0.18\textwidth}
        \centering
        \spyBCN{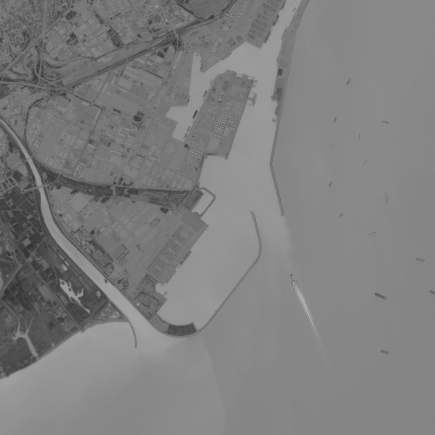}
        \spyBCN{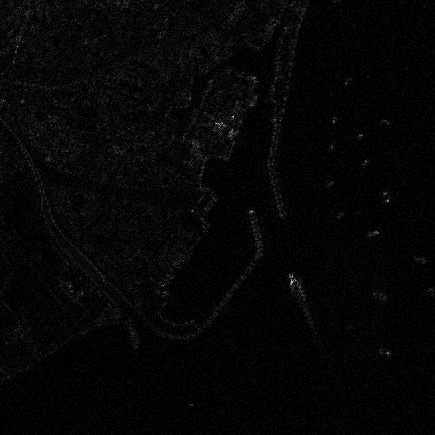}
        \captionsetup{font=small, labelformat=empty}
        \caption{MMNet}
    \end{subfigure}
    \begin{subfigure}[b]{0.18\textwidth}
        \centering
        \spyBCN{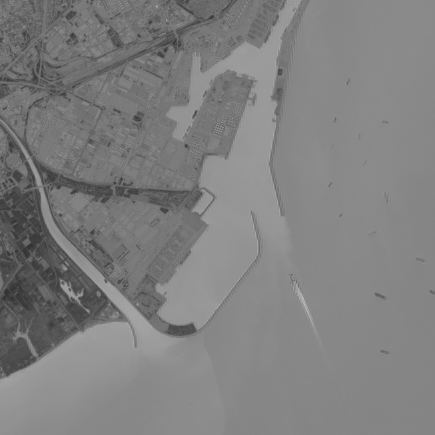}
        \spyBCN{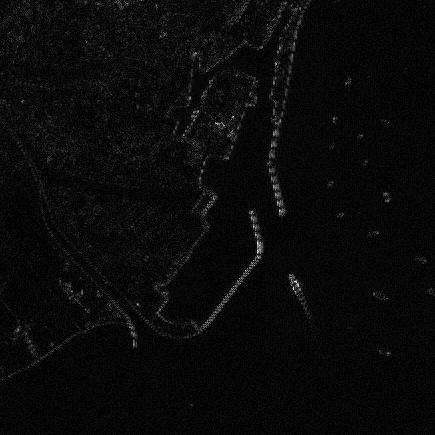}
        \captionsetup{font=small, labelformat=empty}
        \caption{AWFLN}
    \end{subfigure}

    \begin{subfigure}[b]{0.18\textwidth}
        \centering
        \spyBCN{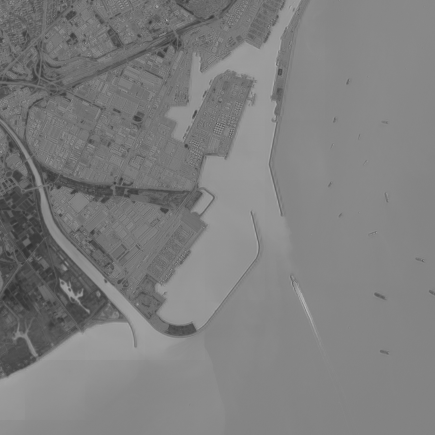}
        \spyBCN{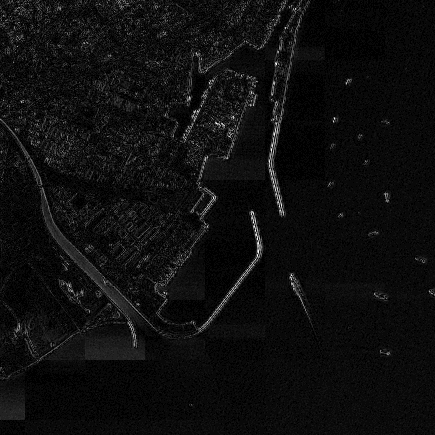}
        \captionsetup{font=small, labelformat=empty}
        \caption{UTeRM}
    \end{subfigure}
    \begin{subfigure}[b]{0.18\textwidth}
        \centering
        \spyBCN{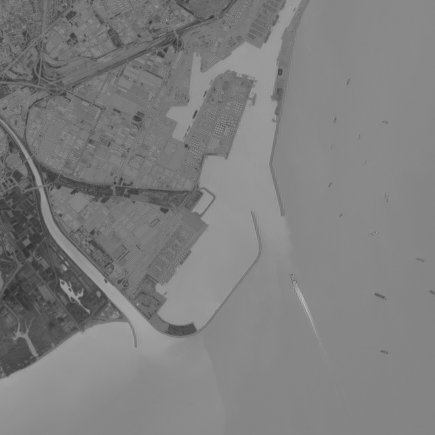}
        \spyBCN{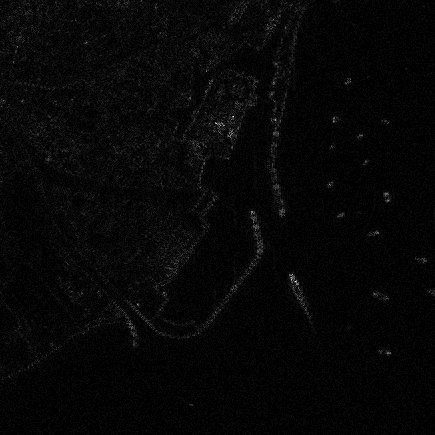}
        \captionsetup{font=small, labelformat=empty}
        \caption{LGCT}
    \end{subfigure}
    \begin{subfigure}[b]{0.18\textwidth}
        \centering
        \spyBCN{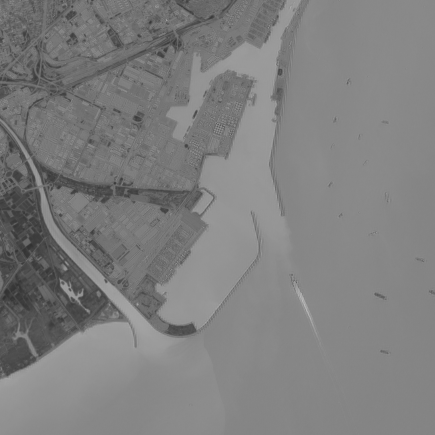}
        \spyBCN{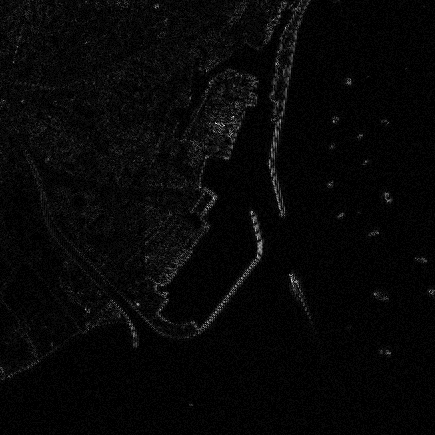}
        \captionsetup{font=small, labelformat=empty}
        \caption{WINet}
    \end{subfigure}
    \begin{subfigure}[b]{0.18\textwidth}
        \centering
        \spyBCN{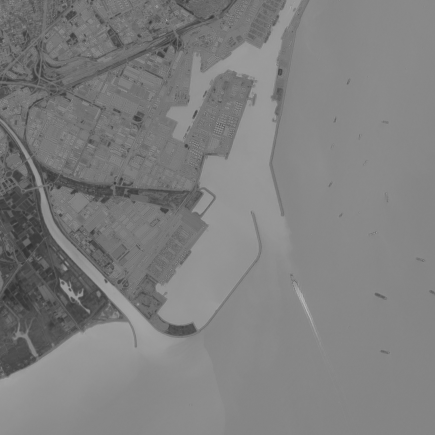}
        \spyBCN{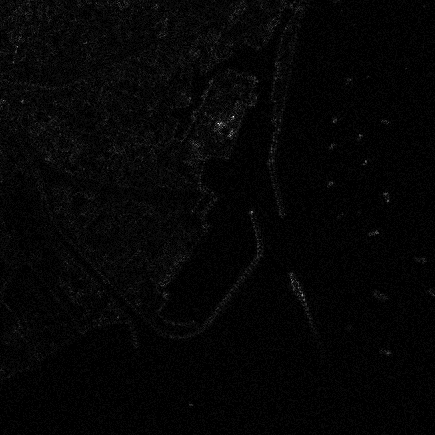}
        \captionsetup{font=small, labelformat=empty}
        \caption{GINet}
    \end{subfigure}
    \begin{subfigure}[b]{0.18\textwidth}
        \centering
        \spyBCN{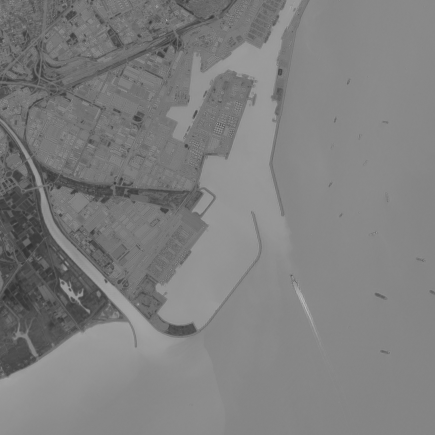}
        \spyBCN{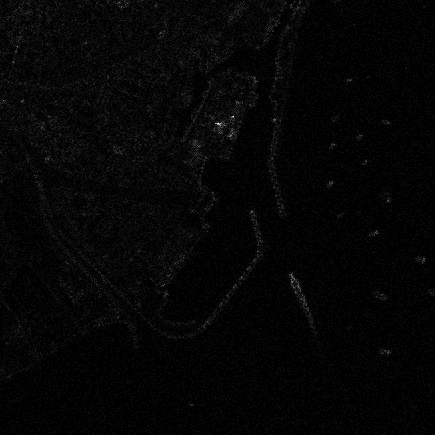}
        \captionsetup{font=small, labelformat=empty}
        \caption{GINet+}
    \end{subfigure}
    \caption[Visual results on the Coastal Sentinel-2 testing set]{Visual comparison on a coastal crop from Barcelona. NDWI and error maps are shown. Most methods produce either over-smoothed outputs or jagged, unnatural edges, as observed along the dock boundary and on the boat in the zoomed-in area. GINet and GINet+ offer the best overall performance.
    }
    \label{fig:barcelona}
\end{figure}


Figure \ref{fig:NovaZelanda} shows the results on the rural crop from New Zealand. In this case, we display the SWIR composites and the corresponding error maps. While Bicubic, CNMF, and UTeRM produce blurred outputs, visual differences among the other methods are difficult to discern. This is mainly due to the use of the 10m B4 band as the blue channel in the composite. To better assess the quality of the results, we refer to the error maps. These reveal that, in addition to the previously mentioned approaches, GSA, FusionNet, AWFLN, ResSen2, and DSen2 introduce higher levels of deviation. Within the other techniques, GINet and GINet+ stand out by yielding the most faithful reconstructions, as can be observed in the zoomed-in areas.

Finally, the results on the coastal crop from Barcelona are presented in Figure \ref{fig:barcelona}. We show the NDWI defined in \eqref{s2-ndwi}, along with the associated error maps. For visualization purposes, the index values have been rescaled between 0 and 1. While Bicubic, CNMF, GSA, and UTeRM produce maps with over-smoothed regions, FusionNet, MMNet, AWFLN, LGCT, and WINet generate jagged and unnatural edges, as can be seen along the dock boundary or on the boat approaching the coast in the zoomed-in area. Among the remaining methods, GINet and GINet+ offer better overall performance, particularly around the ships navigating in open water (see the error maps).

\section{Ablation study} \label{sec:Ablation}

We conduct several experiments to identify the optimal configuration for the proposed guided SR architecture, which is shared by both GINet and GINet+. Specifically, we evaluate alternatives to $\mathcal{R}es_{NL}$, explore different hyperparameter values, and compare various loss function choices. All tests are carried out using the cluster-based guiding image generation strategy.

\begin{table}[t]
\centering
\caption[Quantitative comparison of different architectures for the back-projection kernel in the geometry-guided SR Sentinel-2 model]{Quantitative metrics on the full Sentinel-2 testing set for different architectures of the back-projection kernel. $\mathcal{R}es_{\mathrm{NL}}$ refers to the proposed configuration in Figure~\ref{fig:architecture}. ResNet denotes a residual network with comparable number of parameters, where the 10m and 20m bands are concatenated, while ResNetSR excludes the 10m bands. The best performance, highlighted in bold, is achieve by $\mathcal{R}es_{\mathrm{NL}}$.}
\begin{tabular}{l| c|c|c|c }
 & ERGAS$\downarrow$& PSNR$\uparrow$& SSIM$\uparrow$& SAM$\downarrow$\\
 \hline
$\mathcal{R}es_{\mathrm{NL}}$ & \textbf{0.5546}& \textbf{42.85}& \textbf{0.9809}& \textbf{0.7932} \\
ResNet &  0.5624 & 42.60 & 0.9803 &  0.8047 \\
ResNetSR & 1.1731 & 36.79  & 0.9228 & 1.1157 \\
\hline
\end{tabular}
\label{tab:ablation-arch}
\end{table}

\begin{figure}[t]
    \centering
    \begin{subfigure}[t]{0.24\textwidth}
        \centering
        \spyPalma{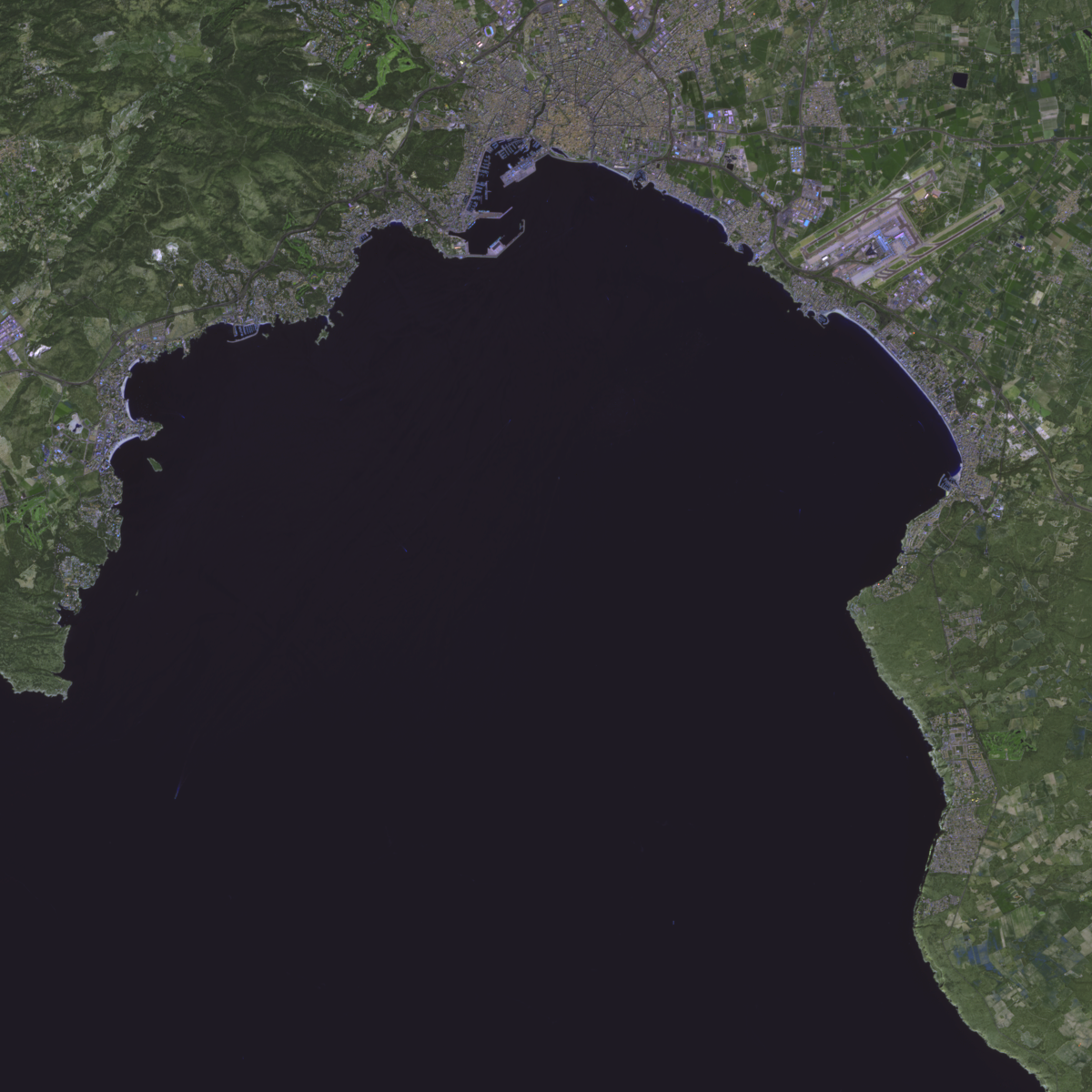}
        \vskip 0.2em
        \spyPalma{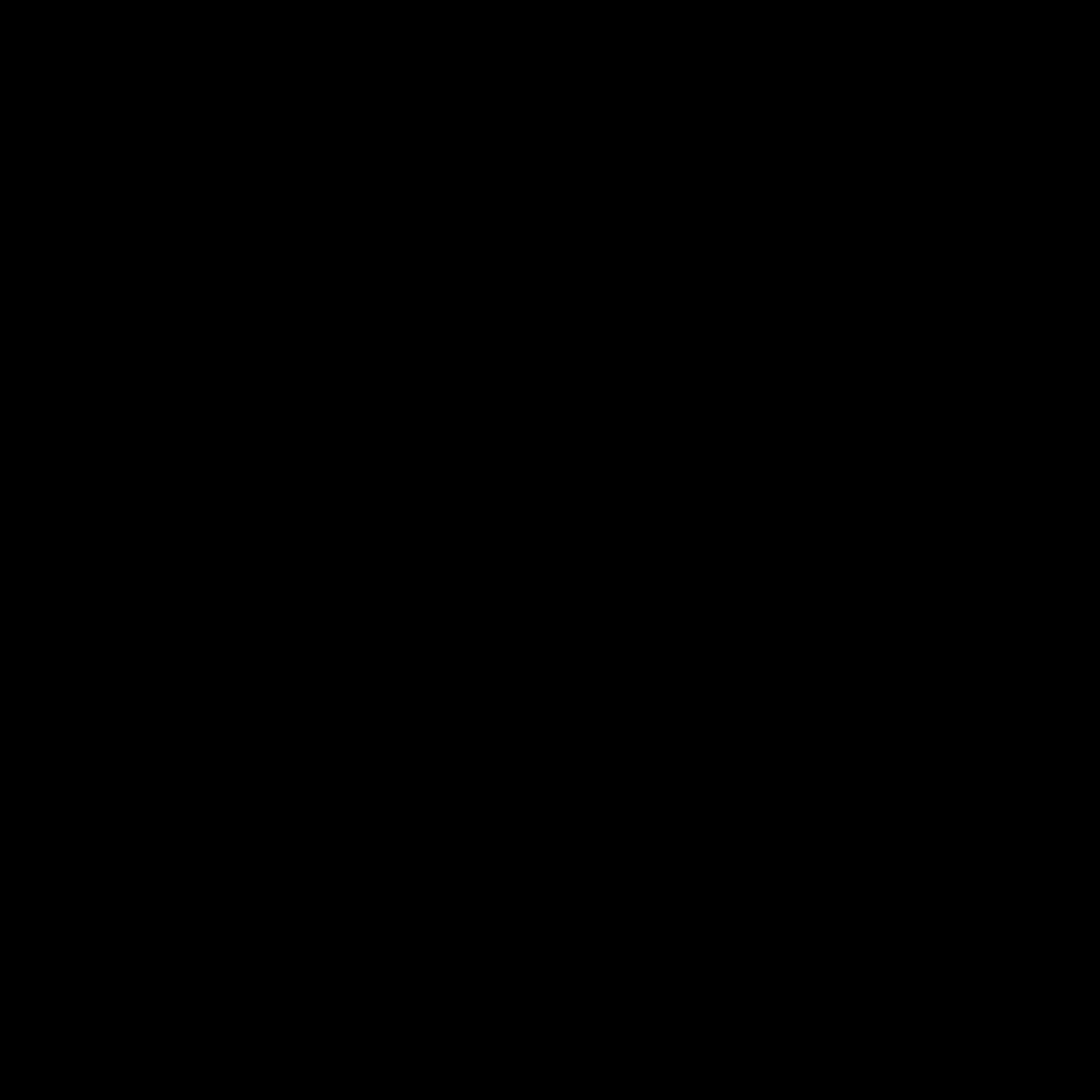}
        \caption*{Reference}
    \end{subfigure}
    \hfill
    \begin{subfigure}[t]{0.24\textwidth}
        \centering
        \spyPalma{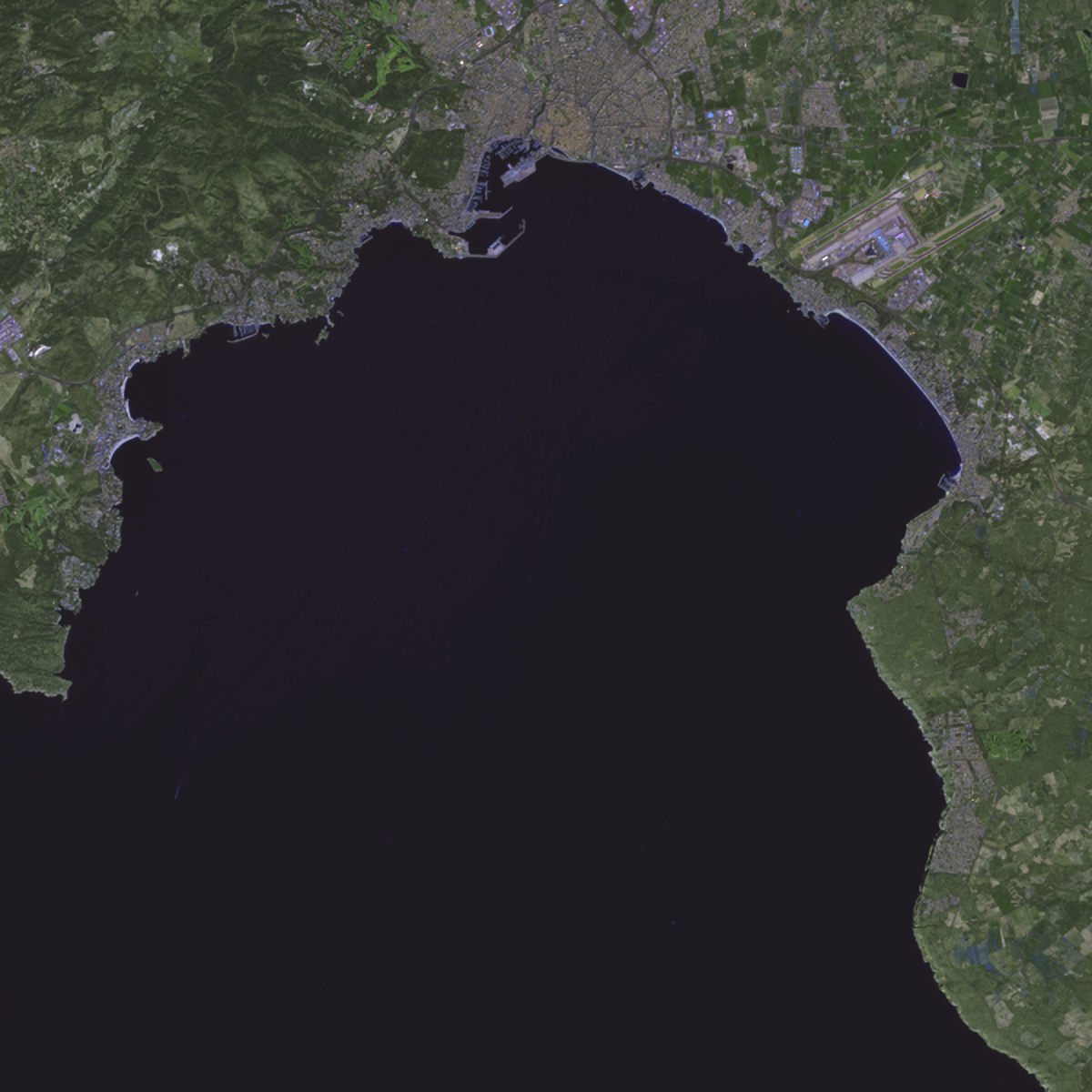}
        \vskip 0.2em
        \spyPalma{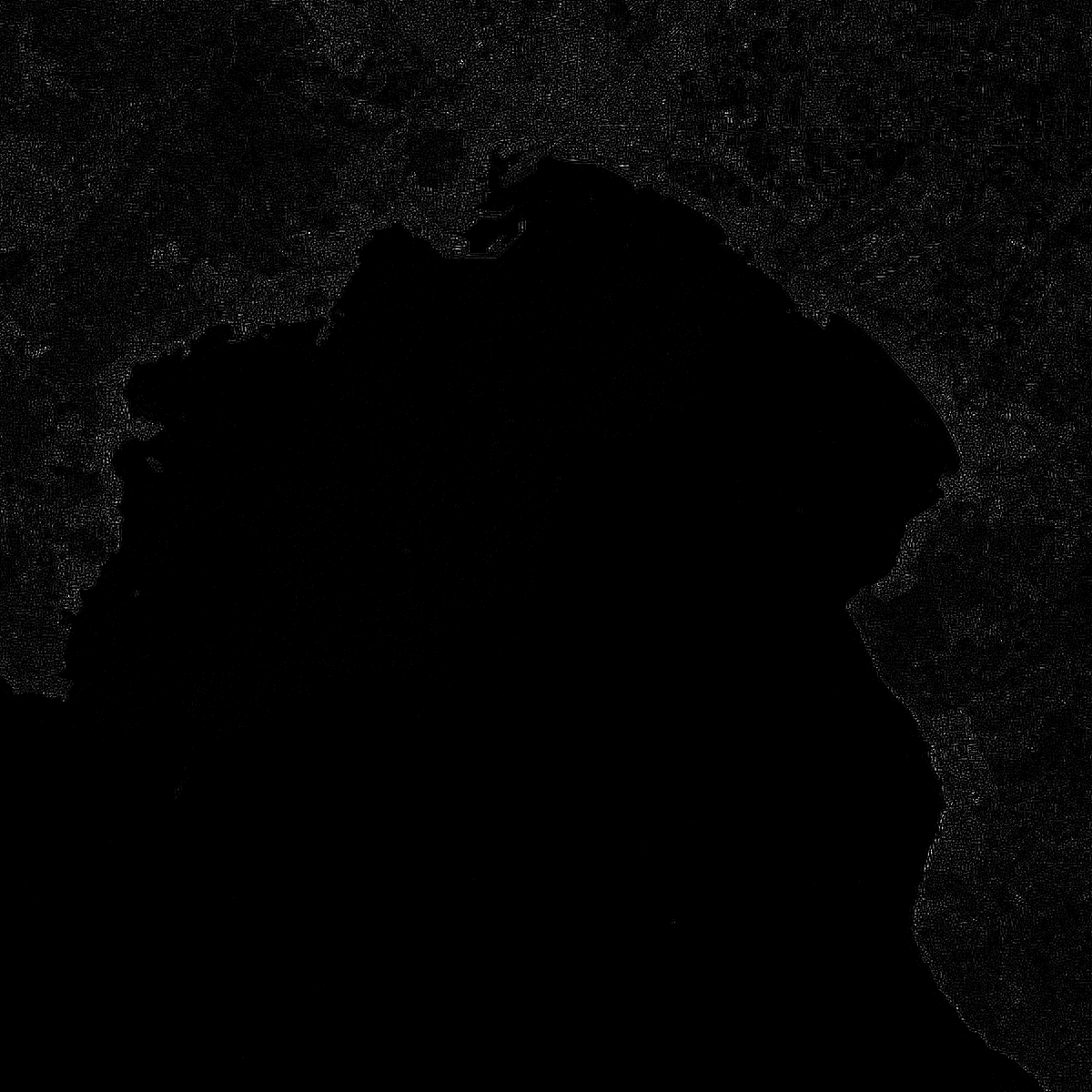}
        \caption*{ResNetSR}
    \end{subfigure}
    \hfill
    \begin{subfigure}[t]{0.24\textwidth}
        \centering
        \spyPalma{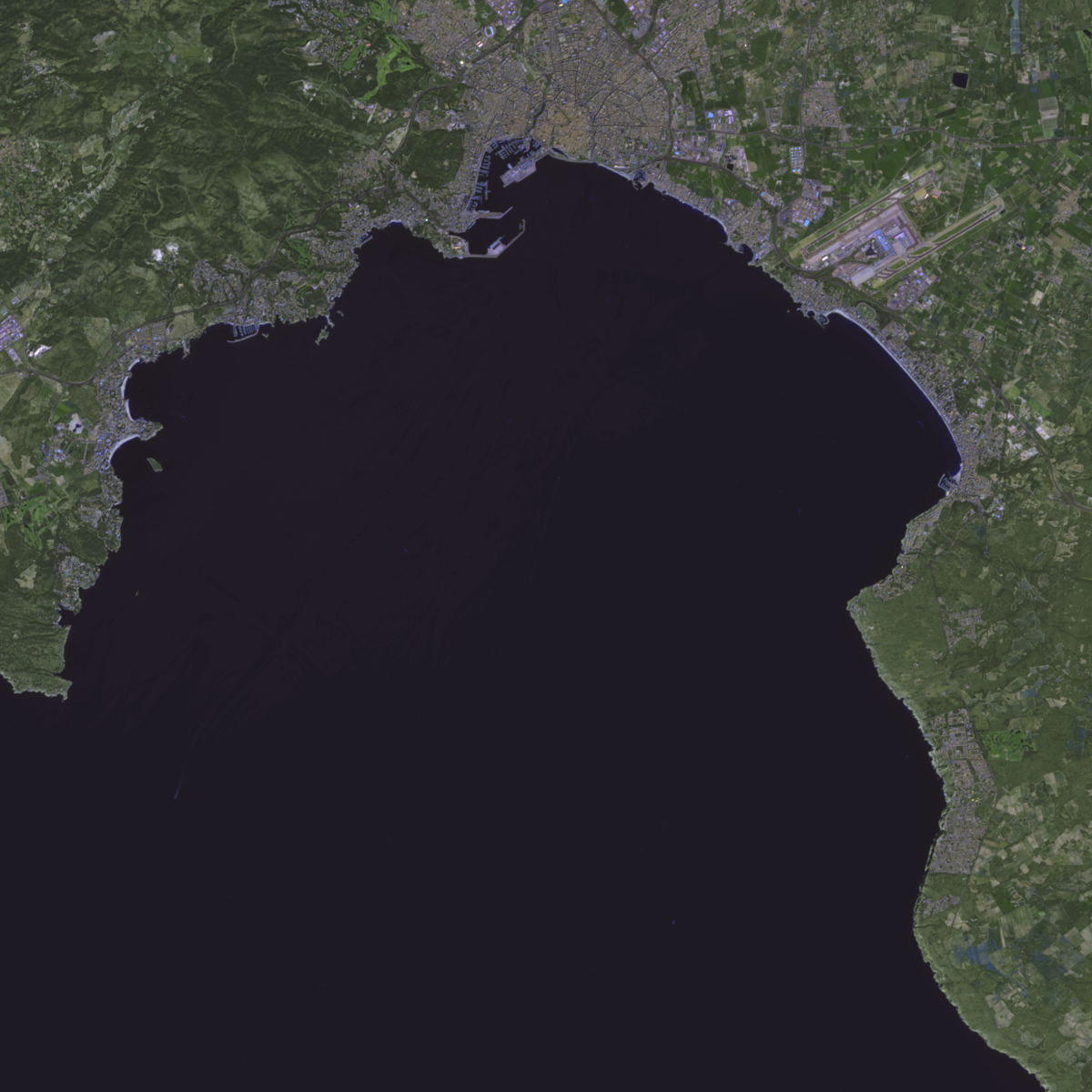}
        \vskip 0.2em
        \spyPalma{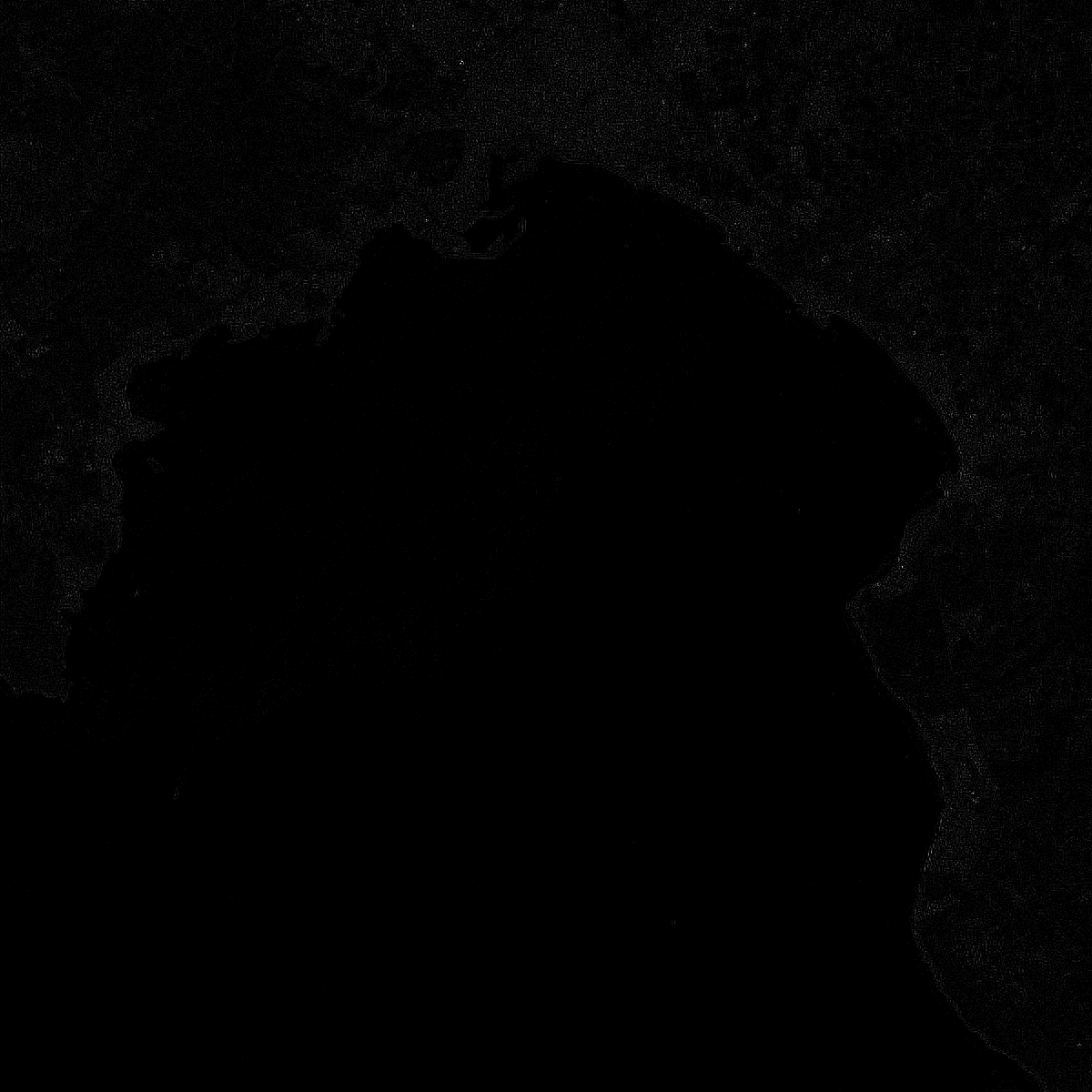}
        \caption*{ResNet}
    \end{subfigure}
    \hfill
    \begin{subfigure}[t]{0.24\textwidth}
        \centering
        \spyPalma{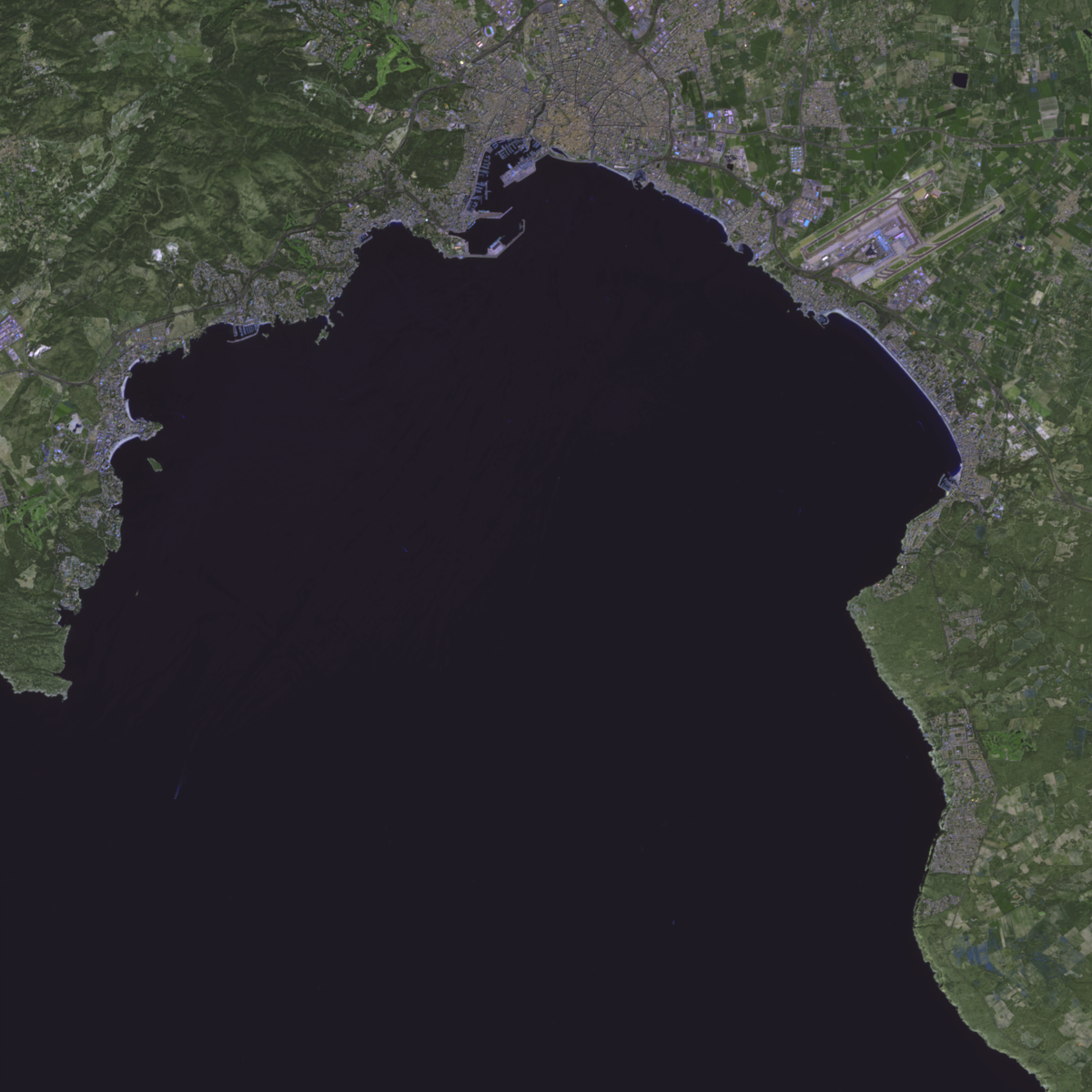}
        \vskip 0.2em
        \spyPalma{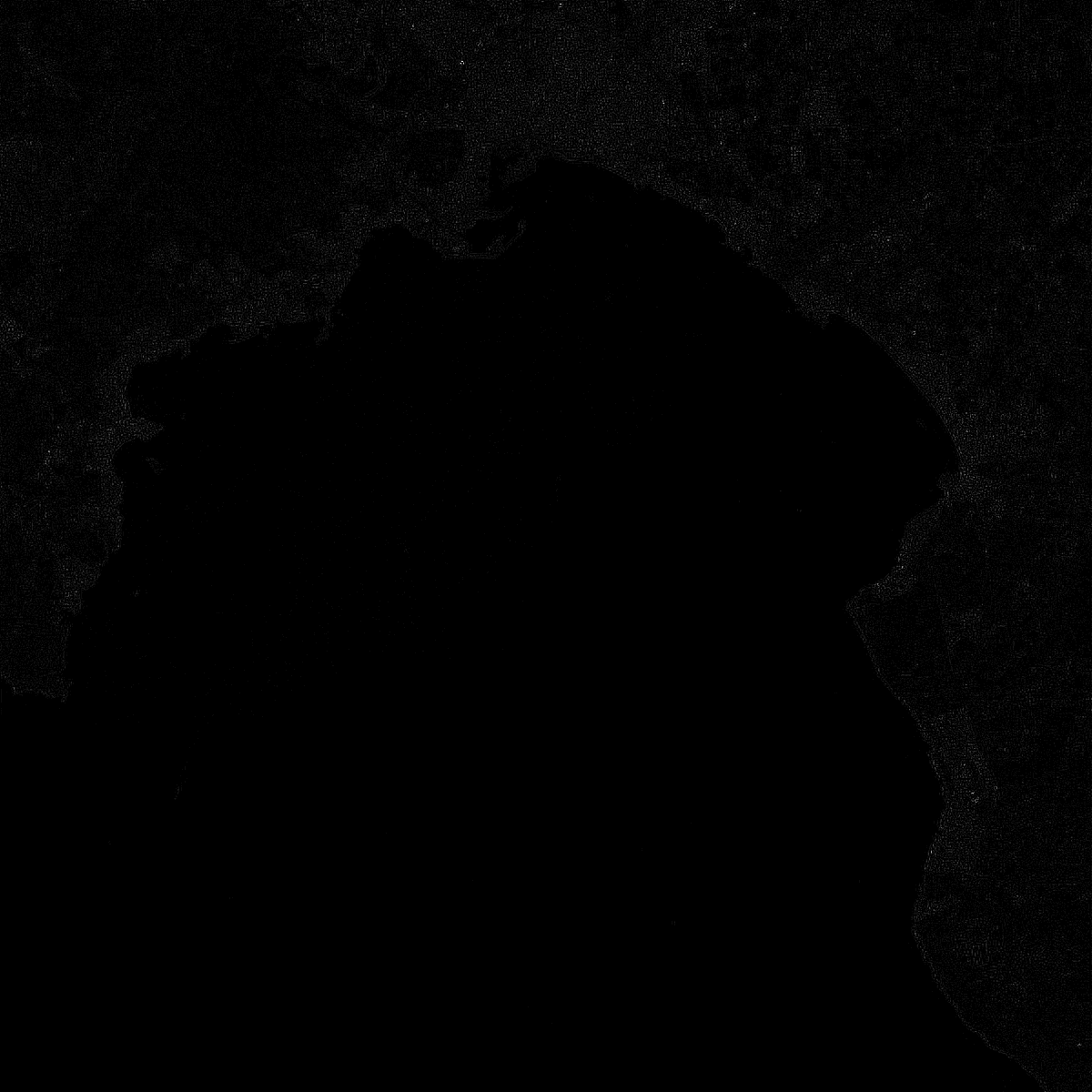}
        \caption*{$\mathcal{R}es_{\mathrm{NL}}$}
    \end{subfigure}
    \caption[Visual comparison of different architectures for the back-projection kernel in the geometry-guided SR Sentinel-2 model]{Visual comparison of different architectures for the back-projection kernel on the Palma testing crop. The top row shows the urban false color composites, while the bottom row displays the error maps. $\mathcal{R}es_{\mathrm{NL}}$ refers to the proposed configuration in Figure~\ref{fig:architecture}. ResNet denotes a residual network with comparable number of parameters, where the 10m and 20m bands are concatenated, while ResNetSR excludes the 10m bands. The results highlight the importance of the guiding image for accurate geometry recovery, and the role of the nonlocal mechanism in reducing artifacts along edges.}    
    \label{fig:ablation}
\end{figure}

Regarding the architecture of the $\mathcal{R}es_{\mathrm{NL}}$ module illustrated in Figure~\ref{fig:architecture}, we compare it with two alternative designs, ensuring a comparable number of learnable parameters in all cases. First, we consider a residual architecture (ResNet) in which the 10m and 20m bands are concatenated, following the same strategy as in the proposed residual module. We also assess the performance of this configuration when the 10m bands are excluded, thereby reformulating the task as a pure SR problem (ResNetSR). The quantitative metrics are reported in Table \ref{tab:ablation-arch}. We observe that removing the attention mechanism leads to a performance drop, while omitting the HR guidance results in a substantial degradation. Figure \ref{fig:ablation} shows the enhanced images produced by each configuration on a crop from Palma, part of the Coastal Sentinel-2 subset. These results highlight the importance of the guiding image for accurately recovering the geometry of the scene, as seen in the zoom-in boxes, and the role of the nonlocal mechanism in reducing artifacts along edges. These observations are further corroborated by the corresponding error maps.

In addition, we study the influence of the number of stages, as well as the patch and window sizes used for computing the attention weights. First, we evaluate the performance with 3, 6, and 9 stages. As illustrated in Table \ref{tab:ablation-stages}, the best PSNR is achieved with 6 stages. Then, keeping the number of stages fixed at 6, we analyse the results obtained using patch and window of sizes $3\times 3$, $5\times 5$, and $7\times 7$. Table \ref{tab:ablation-patch-wind} shows that the optimal performance is attained when the patch and window sizes are set to $3\times 3$ and $5\times 5$, respectively.

\begin{table}[t]
\centering
\caption[Quantitative comparison of different numbers of stages in the geometry-guided SR Sentinel-2 model]{Quantitative metrics on the full Sentinel-2 testing set for varying numbers of stages in the proposed geometry-guided SR model. The best performance, highlighted in bold, is achieved with 6 stages.}
\begin{tabular}{l |c|c|c|c }
 & ERGAS$\downarrow$& PSNR$\uparrow$& SSIM$\uparrow$& SAM$\downarrow$\\
 \hline
3 stages & 0.5649 & 42.59 & 0.9801 & 0.8050 \\
6 stages & \textbf{0.5546}& \textbf{42.85}& \textbf{0.9809}& \textbf{0.7932} \\
9 stages & 0.5705 & 42.41 & 0.9797 & 0.8205 \\
\hline
\end{tabular}
\label{tab:ablation-stages}
\end{table}

\begin{table}[t]
\centering
\caption[Quantitative comparison of different combinations of patch and window sizes in the MHA module of the geometry-guided SR Sentinel-2 model]{PSNR obtained with different combinations of patch and window sizes in the MHA module of the proposed geometry-guided SR model. The best performance, highlighted in bold, is attained when the patch and window sizes are set to $3\times 3$ and $5\times 5$, respectively.}
\begin{tabular}{c| c|c|c}
 & \multicolumn{3}{c}{{Window Size}} \\
\hline
{Patch Size} & 3 & 5 & 7 \\
\hline
3 & 42.51 & 42.85 & 42.40 \\
5 & 42.57 & 42.35 & 42.42 \\
7 & 42.25 & 42.41 & 41.98 \\
\hline
\end{tabular}
\label{tab:ablation-patch-wind}
\end{table}

Finally, we examine the impact of the loss function while keeping all hyperparameters fixed. In addition to the classical L1 and MSE losses computed between the output and the ground truth, $\Uref$, we explore a weighted MSE over all intermediate stages combined with the L1 loss as follows:
\begin{equation}
\mathcal{L}_{i,2}^{\alpha} = \|\Uref-u^K\|^i_{i,i} + \alpha\sum_{k=1}^{K-1}\|\Uref-u^k\|_{2,2}.   
\end{equation}
Note that the L1 and MSE losses correspond to $\mathcal{L}_{1,2}^0$ and $\mathcal{L}_{2,2}^0$, respectively. We test this approach with $\alpha=0.1$ and $\alpha=0.5$. As shown in Table \ref{tab:ablation-loss}, the best PSNR is achieved  using only the L1 loss between the output and the ground truth.

\begin{table}[t]
\centering
\caption[Quantitative comparison of different loss functions for the Sentinel-2 SR method]{Quantitative metrics on the full Sentinel-2 testing set for different loss functions. The best performance, highlighted in bold, is attained when using the L1 loss between the output and the ground truth image.}
\begin{tabular}{l |c|c|c|c }
 & ERGAS$\downarrow$& PSNR$\uparrow$& SSIM$\uparrow$& SAM$\downarrow$\\
 \hline
L1 & \textbf{0.5546}& \textbf{42.85}& \textbf{0.9809}& \textbf{0.7932} \\
MSE & 0.5626 & 42.55 & 0.9797 & 0.8063  \\
$\mathcal{L}_{1,2}^{\alpha},$ $\alpha=0.1$ & 0.5631 & 42.65 & 0.9804 & 0.8061 \\
$\mathcal{L}_{1,2}^{\alpha},$ $\alpha=0.5$ & 0.5751 & 42.48 & 0.9795 & 0.8258 \\
\hline
\end{tabular}
\label{tab:ablation-loss}
\end{table}

\section{Conclusions} \label{sec:Conclusion}

In this paper, we have developed a geometry-guided super-resolution model based on an unfolded back-projection framework for the fusion of 10m and 20m bands acquired by the Sentinel-2 mission. Our approach leverages the spatial detail present in a generated guiding image and exploits image self-similarities to enhance the spatial resolution of the 20m bands. To construct the guiding image, we have introduced a novel cluster-based learning procedure that captures the geometric information encoded in the 10m bands. This geometry-rich representation effectively drives the super-resolution process, which is formulated as an unfolded back-projection algorithm. To avoid artifacts typically associated with back-projection, the classical kernel has been replaced by a residual network incorporating a MHA module, which models patch-based nonlocal interactions across spatial and spectral dimensions.

Our methods have been trained using a dataset of Sentinel-2 crops from six geographical areas and evaluated on three testing sets featuring urban, rural, and coastal landscapes. Comparative results demonstrate that GINet and GINet+ consistently outperform both classical and deep learning-based fusion and super-resolution techniques, including those tailored specifically for Sentinel-2 data. The relative performance of competing methods varies depending on the landscape, while our approaches deliver superior results across all scenarios, highlighting the effectiveness of their hybrid attention mechanisms and model-based formulation in adapting to the diverse spatial and spectral characteristics inherent in Sentinel-2 imagery.

By evaluating the proposed model using a variety of color composites and indices commonly employed in remote sensing, we demonstrate its effectiveness in enhancing the visual and analytical quality of Sentinel-2 imagery. This improvement could contribute directly to more accurate detection and interpretation of diverse Earth surface phenomena, thereby reinforcing the utility of the model in practical remote sensing applications.

Our study has concentrated exclusively on 10m and 20m bands, without addressing 60m bands. Nonetheless, these lower-resolution bands contain valuable spectral information that can be crucial for a variety of applications, such as atmospheric correction and water quality assessment. Consequently, future research could focus on extending the proposed model to super-resolve the 60m bands by leveraging the spatial details present in the 10m and/or 20m bands. Such an extension would enable a more comprehensive and consistent high-resolution product across all Sentinel-2 spectral ranges.

Moreover, given that our framework is designed to jointly process multiple spectral bands at varying spatial resolutions, we hypothesize that the methodology can be generalized beyond Sentinel-2 data. Specifically, it could be adapted to enhance the spatial resolution of other multispectral or hyperspectral satellite platforms, potentially improving the quality and utility of diverse Earth observation datasets.

On the other hand, we have incorporated a MHA module that has proven effective in taking advantage of self-similarities in satellite images. However, this module introduces significant computational overhead, which may limit its practical deployment on resource-constrained hardware. Therefore, future work could also focus on optimizing the current MHA module or replacing it with a more computationally efficient architecture that still maintains the capability to exploit nonlocal interactions. Such improvements would greatly enhance the scalability and applicability of the model in real-world operational settings.

\section*{Acknowledgements}
\sloppy This work was funded by the European Union NextGenerationEU/PRTR via MaLiSat project TED2021-132644B-I00, and also by MCIN/AEI/10.13039/501100011033 and
“ERDF A way of making Europe” through European Union under Grant PID2021-125711OB-I00. Ivan Pereira-Sánchez is grateful for the funding provided by the Conselleria de Fons Europeus, Universitat i Cultura (GOIB) under grant FPU2023-004-C. Daniel Torres is grateful for the funding provided by the Conselleria d'Educació i Universitats (GOIB) under grant FPU2024-002-C. Bartomeu Garau is grateful for the funding provided by the Conselleria d'Educació i Universitats (GOIB) under grant FPU2025-008-C. The authors gratefully acknowledge the computer resources at Artemisa, funded by the EU ERDF and Comunitat Valenciana and the technical support provided by IFIC (CSIC-UV).

\bibliographystyle{elsarticle-harv} 
\bibliography{bibliography}

\begin{thebibliography}{77}
\expandafter\ifx\csname natexlab\endcsname\relax\def\natexlab#1{#1}\fi
\providecommand{\url}[1]{\texttt{#1}}
\providecommand{\href}[2]{#2}
\providecommand{\path}[1]{#1}
\providecommand{\DOIprefix}{doi:}
\providecommand{\ArXivprefix}{arXiv:}
\providecommand{\URLprefix}{URL: }
\providecommand{\Pubmedprefix}{pmid:}
\providecommand{\doi}[1]{\href{http://dx.doi.org/#1}{\path{#1}}}
\providecommand{\Pubmed}[1]{\href{pmid:#1}{\path{#1}}}
\providecommand{\bibinfo}[2]{#2}
\ifx\xfnm\relax \def\xfnm[#1]{\unskip,\space#1}\fi
\bibitem[{Addabbo et~al.(2016)Addabbo, Focareta, Marcuccio, Votto, Ullo et~al.}]{2016-vegetation}
\bibinfo{author}{Addabbo, P.}, \bibinfo{author}{Focareta, M.}, \bibinfo{author}{Marcuccio, S.}, \bibinfo{author}{Votto, C.}, \bibinfo{author}{Ullo, S.L.}, et~al., \bibinfo{year}{2016}.
\newblock \bibinfo{title}{Contribution of sentinel-2 data for applications in vegetation monitoring}.
\newblock \bibinfo{journal}{Acta Imeko} \bibinfo{volume}{5}, \bibinfo{pages}{44--54}.
\bibitem[{Aiazzi et~al.(2003)Aiazzi, Alparone, Baronti, Garzelli and Selva}]{aiazzi2003mtf}
\bibinfo{author}{Aiazzi, B.}, \bibinfo{author}{Alparone, L.}, \bibinfo{author}{Baronti, S.}, \bibinfo{author}{Garzelli, A.}, \bibinfo{author}{Selva, M.}, \bibinfo{year}{2003}.
\newblock \bibinfo{title}{An mtf-based spectral distortion minimizing model for pan-sharpening of very high resolution multispectral images of urban areas}, in: \bibinfo{booktitle}{2003 2nd GRSS/ISPRS Joint Workshop on Remote Sensing and Data Fusion over Urban Areas}, \bibinfo{organization}{IEEE}. pp. \bibinfo{pages}{90--94}.
\bibitem[{Aldo{\u{g}}an et~al.(2024)Aldo{\u{g}}an, Aksu and Demirel}]{aldougan2024enhancement}
\bibinfo{author}{Aldo{\u{g}}an, C.}, \bibinfo{author}{Aksu, K.}, \bibinfo{author}{Demirel, H.}, \bibinfo{year}{2024}.
\newblock \bibinfo{title}{Enhancement of sentinel-2a images for ship detection via real-esrgan model}.
\newblock \bibinfo{journal}{Applied Sciences} \bibinfo{volume}{14}, \bibinfo{pages}{11988}.
\bibitem[{Alparone et~al.(2024)Alparone, Arienzo and Garzelli}]{SpatialResEnhanc2024}
\bibinfo{author}{Alparone, L.}, \bibinfo{author}{Arienzo, A.}, \bibinfo{author}{Garzelli, A.}, \bibinfo{year}{2024}.
\newblock \bibinfo{title}{Spatial resolution enhancement of satellite hyperspectral data via nested hypersharpening with sentinel-2 multispectral data}.
\newblock \bibinfo{journal}{IEEE Journal of Selected Topics in Applied Earth Observations and Remote Sensing} \bibinfo{volume}{17}, \bibinfo{pages}{10956--10966}.
\bibitem[{Alparone et~al.(2004)Alparone, Baronti, Garzelli and Nencini}]{alparone2004global}
\bibinfo{author}{Alparone, L.}, \bibinfo{author}{Baronti, S.}, \bibinfo{author}{Garzelli, A.}, \bibinfo{author}{Nencini, F.}, \bibinfo{year}{2004}.
\newblock \bibinfo{title}{A global quality measurement of pan-sharpened multispectral imagery}.
\newblock \bibinfo{journal}{IEEE Geoscience and Remote Sensing Letters} \bibinfo{volume}{1}, \bibinfo{pages}{313--317}.
\bibitem[{Ballester et~al.(2006)Ballester, Caselles, Igual, Verdera and Roug{\'e}}]{ballester2006variational}
\bibinfo{author}{Ballester, C.}, \bibinfo{author}{Caselles, V.}, \bibinfo{author}{Igual, L.}, \bibinfo{author}{Verdera, J.}, \bibinfo{author}{Roug{\'e}, B.}, \bibinfo{year}{2006}.
\newblock \bibinfo{title}{A variational model for p+ xs image fusion}.
\newblock \bibinfo{journal}{International Journal of Computer Vision} \bibinfo{volume}{69}, \bibinfo{pages}{43--58}.
\bibitem[{Cai and Huang(2021)}]{cai2020super}
\bibinfo{author}{Cai, J.}, \bibinfo{author}{Huang, B.}, \bibinfo{year}{2021}.
\newblock \bibinfo{title}{Super-resolution-guided progressive pansharpening based on a deep convolutional neural network}.
\newblock \bibinfo{journal}{IEEE Transactions on Geoscience and Remote Sensing} \bibinfo{volume}{59}, \bibinfo{pages}{5206--5220}.
\bibitem[{Dai et~al.(2007)Dai, Han, Wu and Gong}]{dai2007bilateral}
\bibinfo{author}{Dai, S.}, \bibinfo{author}{Han, M.}, \bibinfo{author}{Wu, Y.}, \bibinfo{author}{Gong, Y.}, \bibinfo{year}{2007}.
\newblock \bibinfo{title}{Bilateral back-projection for single image super resolution}, in: \bibinfo{booktitle}{2007 IEEE International Conference on Multimedia and Expo}, pp. \bibinfo{pages}{1039--1042}.
\bibitem[{Deng et~al.(2020)Deng, Vivone, Jin and Chanussot}]{deng2020detail}
\bibinfo{author}{Deng, L.J.}, \bibinfo{author}{Vivone, G.}, \bibinfo{author}{Jin, C.}, \bibinfo{author}{Chanussot, J.}, \bibinfo{year}{2020}.
\newblock \bibinfo{title}{Detail injection-based deep convolutional neural networks for pansharpening}.
\newblock \bibinfo{journal}{IEEE Transactions on Geoscience and Remote Sensing} \bibinfo{volume}{59}, \bibinfo{pages}{6995--7010}.
\bibitem[{Deudon et~al.(2020)Deudon, Kalaitzis, Goytom, Arefin, Lin, Sankaran, Michalski, Kahou, Cornebise and Bengio}]{deudon2020highres}
\bibinfo{author}{Deudon, M.}, \bibinfo{author}{Kalaitzis, A.}, \bibinfo{author}{Goytom, I.}, \bibinfo{author}{Arefin, M.}, \bibinfo{author}{Lin, Z.}, \bibinfo{author}{Sankaran, K.}, \bibinfo{author}{Michalski, V.}, \bibinfo{author}{Kahou, S.}, \bibinfo{author}{Cornebise, J.}, \bibinfo{author}{Bengio, Y.}, \bibinfo{year}{2020}.
\newblock \bibinfo{title}{Highres-net: Recursive fusion for multi-frame super-resolution of satellite imagery}.
\newblock \bibinfo{journal}{arXiv preprint arXiv:2002.06460} .
\bibitem[{Dong et~al.(2014)Dong, Loy, He and Tang}]{dong2014learning}
\bibinfo{author}{Dong, C.}, \bibinfo{author}{Loy, C.}, \bibinfo{author}{He, K.}, \bibinfo{author}{Tang, X.}, \bibinfo{year}{2014}.
\newblock \bibinfo{title}{Learning a deep convolutional network for image super-resolution}, in: \bibinfo{booktitle}{Computer Vision--ECCV 2014: 13th European Conference, Zurich, Switzerland, September 6-12, 2014, Proceedings, Part IV 13}, \bibinfo{organization}{Springer}. pp. \bibinfo{pages}{184--199}.
\bibitem[{Dong et~al.(2009)Dong, Zhang, Shi and Wu}]{dong2009nonlocal}
\bibinfo{author}{Dong, W.}, \bibinfo{author}{Zhang, L.}, \bibinfo{author}{Shi, G.}, \bibinfo{author}{Wu, X.}, \bibinfo{year}{2009}.
\newblock \bibinfo{title}{Nonlocal back-projection for adaptive image enlargement}, in: \bibinfo{booktitle}{2009 16th IEEE International Conference on Image Processing (ICIP)}, pp. \bibinfo{pages}{349--352}.
\bibitem[{Dosovitskiy et~al.(2020)Dosovitskiy, Beyer, Kolesnikov, Weissenborn, Zhai, Unterthiner, Dehghani, Minderer, Heigold, Gelly et~al.}]{dosovitskiy2020image}
\bibinfo{author}{Dosovitskiy, A.}, \bibinfo{author}{Beyer, L.}, \bibinfo{author}{Kolesnikov, A.}, \bibinfo{author}{Weissenborn, D.}, \bibinfo{author}{Zhai, X.}, \bibinfo{author}{Unterthiner, T.}, \bibinfo{author}{Dehghani, M.}, \bibinfo{author}{Minderer, M.}, \bibinfo{author}{Heigold, G.}, \bibinfo{author}{Gelly, S.}, et~al., \bibinfo{year}{2020}.
\newblock \bibinfo{title}{An image is worth 16x16 words: Transformers for image recognition at scale}.
\newblock \bibinfo{journal}{arXiv preprint arXiv:2010.11929} .
\bibitem[{Du et~al.(2016)Du, Zhang, Ling, Wang, Li and Li}]{2016-water}
\bibinfo{author}{Du, Y.}, \bibinfo{author}{Zhang, Y.}, \bibinfo{author}{Ling, F.}, \bibinfo{author}{Wang, Q.}, \bibinfo{author}{Li, W.}, \bibinfo{author}{Li, X.}, \bibinfo{year}{2016}.
\newblock \bibinfo{title}{Water bodies’ mapping from sentinel-2 imagery with modified normalized difference water index at 10-m spatial resolution produced by sharpening the swir band}.
\newblock \bibinfo{journal}{Remote Sensing} \bibinfo{volume}{8}, \bibinfo{pages}{354}.
\bibitem[{Duran et~al.(2014)Duran, Buades, Coll and Sbert}]{duran2014nonlocal}
\bibinfo{author}{Duran, J.}, \bibinfo{author}{Buades, A.}, \bibinfo{author}{Coll, B.}, \bibinfo{author}{Sbert, C.}, \bibinfo{year}{2014}.
\newblock \bibinfo{title}{A nonlocal variational model for pansharpening image fusion}.
\newblock \bibinfo{journal}{SIAM Journal on Imaging Sciences} \bibinfo{volume}{7}, \bibinfo{pages}{761--796}.
\bibitem[{{European Space Agency}()}]{sentinel2-wiki}
\bibinfo{author}{{European Space Agency}}, .
\newblock \bibinfo{title}{Sentinel-2 — sentiwiki}.
\newblock \bibinfo{howpublished}{\url{https://sentiwiki.copernicus.eu/web/sentinel-2}}.
\newblock \bibinfo{note}{Accessed: 2025-05-16}.
\bibitem[{Fernandez-Beltran et~al.(2018)Fernandez-Beltran, Haut, Paoletti, Plaza, Plaza and Pla}]{fernandez2018multimodal}
\bibinfo{author}{Fernandez-Beltran, R.}, \bibinfo{author}{Haut, J.}, \bibinfo{author}{Paoletti, M.}, \bibinfo{author}{Plaza, J.}, \bibinfo{author}{Plaza, A.}, \bibinfo{author}{Pla, F.}, \bibinfo{year}{2018}.
\newblock \bibinfo{title}{Multimodal probabilistic latent semantic analysis for sentinel-1 and sentinel-2 image fusion}.
\newblock \bibinfo{journal}{IEEE Geoscience and Remote Sensing Letters} \bibinfo{volume}{15}, \bibinfo{pages}{1347--1351}.
\bibitem[{Galar et~al.(2019)Galar, Sesma, Ayala and Aranda}]{galar2019super}
\bibinfo{author}{Galar, M.}, \bibinfo{author}{Sesma, R.}, \bibinfo{author}{Ayala, C.}, \bibinfo{author}{Aranda, C.}, \bibinfo{year}{2019}.
\newblock \bibinfo{title}{Super-resolution for sentinel-2 images}.
\newblock \bibinfo{journal}{The International Archives of the Photogrammetry, Remote Sensing and Spatial Information Sciences} \bibinfo{volume}{42}, \bibinfo{pages}{95--102}.
\bibitem[{Gao(1996)}]{GAO1996NDWI}
\bibinfo{author}{Gao, B.}, \bibinfo{year}{1996}.
\newblock \bibinfo{title}{Ndwi—a normalized difference water index for remote sensing of vegetation liquid water from space}.
\newblock \bibinfo{journal}{Remote Sensing of Environment} \bibinfo{volume}{58}, \bibinfo{pages}{257--266}.
\bibitem[{Gascon et~al.(2017)Gascon, Bouzinac, Th{\'e}paut, Jung, Francesconi, Louis, Lonjou, Lafrance, Massera, Gaudel-Vacaresse et~al.}]{2017-construction}
\bibinfo{author}{Gascon, F.}, \bibinfo{author}{Bouzinac, C.}, \bibinfo{author}{Th{\'e}paut, O.}, \bibinfo{author}{Jung, M.}, \bibinfo{author}{Francesconi, B.}, \bibinfo{author}{Louis, J.}, \bibinfo{author}{Lonjou, V.}, \bibinfo{author}{Lafrance, B.}, \bibinfo{author}{Massera, S.}, \bibinfo{author}{Gaudel-Vacaresse, A.}, et~al., \bibinfo{year}{2017}.
\newblock \bibinfo{title}{Copernicus sentinel-2a calibration and products validation status}.
\newblock \bibinfo{journal}{Remote Sensing} \bibinfo{volume}{9}, \bibinfo{pages}{584}.
\bibitem[{Ga{\v{s}}parovi{\'c} and Jogun(2018)}]{gavsparovic2018effect}
\bibinfo{author}{Ga{\v{s}}parovi{\'c}, M.}, \bibinfo{author}{Jogun, T.}, \bibinfo{year}{2018}.
\newblock \bibinfo{title}{The effect of fusing sentinel-2 bands on land-cover classification}.
\newblock \bibinfo{journal}{International journal of remote sensing} \bibinfo{volume}{39}, \bibinfo{pages}{822--841}.
\bibitem[{Gillespie et~al.(1987)Gillespie, Kahle and Walker}]{gillespie1987color}
\bibinfo{author}{Gillespie, A.}, \bibinfo{author}{Kahle, A.}, \bibinfo{author}{Walker, R.}, \bibinfo{year}{1987}.
\newblock \bibinfo{title}{Color enhancement of highly correlated images. ii. channel ratio and “chromaticity” transformation techniques}.
\newblock \bibinfo{journal}{Remote Sensing of Environment} \bibinfo{volume}{22}, \bibinfo{pages}{343--365}.
\bibitem[{Hallada and Cox(1983)}]{hallada1983image}
\bibinfo{author}{Hallada, W.}, \bibinfo{author}{Cox, S.}, \bibinfo{year}{1983}.
\newblock \bibinfo{title}{Image sharpening for mixed spatial and spectral resolution satellite systems}, in: \bibinfo{booktitle}{International Symposium on Remote Sensing of Environment}, pp.~\bibinfo{pages}{--}.
\bibitem[{Haris et~al.(2018)Haris, Shakhnarovich and Ukita}]{haris2018deep}
\bibinfo{author}{Haris, M.}, \bibinfo{author}{Shakhnarovich, G.}, \bibinfo{author}{Ukita, N.}, \bibinfo{year}{2018}.
\newblock \bibinfo{title}{Deep back-projection networks for super-resolution}, in: \bibinfo{booktitle}{Proceedings of the IEEE conference on computer vision and pattern recognition}, pp. \bibinfo{pages}{1664--1673}.
\bibitem[{He et~al.(2024)He, Fu, Li, Ren and Jia}]{2024LGCT}
\bibinfo{author}{He, W.}, \bibinfo{author}{Fu, X.}, \bibinfo{author}{Li, N.}, \bibinfo{author}{Ren, Q.}, \bibinfo{author}{Jia, S.}, \bibinfo{year}{2024}.
\newblock \bibinfo{title}{Lgct: Local-global collaborative transformer for fusion of hyperspectral and multispectral images}.
\newblock \bibinfo{journal}{IEEE Transactions on Geoscience and Remote Sensing} \bibinfo{volume}{62}, \bibinfo{pages}{1--14}.
\bibitem[{Irani and Peleg(1991)}]{Irani1991BackProj}
\bibinfo{author}{Irani, M.}, \bibinfo{author}{Peleg, S.}, \bibinfo{year}{1991}.
\newblock \bibinfo{title}{Improving resolution by image registration}.
\newblock \bibinfo{journal}{CVGIP: Graphical Models and Image Processing} \bibinfo{volume}{53}, \bibinfo{pages}{231--239}.
\bibitem[{Irani and Peleg(1993)}]{irani1993motion}
\bibinfo{author}{Irani, M.}, \bibinfo{author}{Peleg, S.}, \bibinfo{year}{1993}.
\newblock \bibinfo{title}{Motion analysis for image enhancement: Resolution, occlusion, and transparency}.
\newblock \bibinfo{journal}{Journal of Visual Communication and Image Representation} \bibinfo{volume}{4}, \bibinfo{pages}{324--335}.
\bibitem[{Kaplan(2018)}]{S2PanCompar2018}
\bibinfo{author}{Kaplan, G.}, \bibinfo{year}{2018}.
\newblock \bibinfo{title}{Sentinel-2 pan sharpening—comparative analysis}.
\newblock \bibinfo{journal}{Proceedings} \bibinfo{volume}{2}.
\bibitem[{Kato et~al.(2021)Kato, Miyamoto, Amici, Oda, Matsushita and Nakamura}]{kato2021swir}
\bibinfo{author}{Kato, S.}, \bibinfo{author}{Miyamoto, H.}, \bibinfo{author}{Amici, S.}, \bibinfo{author}{Oda, A.}, \bibinfo{author}{Matsushita, H.}, \bibinfo{author}{Nakamura, R.}, \bibinfo{year}{2021}.
\newblock \bibinfo{title}{Automated classification of heat sources detected using swir remote sensing}.
\newblock \bibinfo{journal}{International Journal of Applied Earth Observation and Geoinformation} \bibinfo{volume}{103}, \bibinfo{pages}{102491}.
\bibitem[{Khan et~al.(2008)Khan, Chanussot, Condat and Montanvert}]{khan2008indusion}
\bibinfo{author}{Khan, M.}, \bibinfo{author}{Chanussot, J.}, \bibinfo{author}{Condat, L.}, \bibinfo{author}{Montanvert, A.}, \bibinfo{year}{2008}.
\newblock \bibinfo{title}{Indusion: Fusion of multispectral and panchromatic images using the induction scaling technique}.
\newblock \bibinfo{journal}{IEEE Geoscience and Remote Sensing Letters} \bibinfo{volume}{5}, \bibinfo{pages}{98--102}.
\bibitem[{Kingma(2014)}]{kingma2014adam}
\bibinfo{author}{Kingma, D.P.}, \bibinfo{year}{2014}.
\newblock \bibinfo{title}{Adam: A method for stochastic optimization}.
\newblock \bibinfo{journal}{arXiv preprint arXiv:1412.6980} .
\bibitem[{Kremezi et~al.(2022)Kremezi, Kristollari, Karathanassi, Topouzelis, Kolokoussis, Taggio, Aiello, Ceriola, Barbone and Corradi}]{worldview-sentinel2}
\bibinfo{author}{Kremezi, M.}, \bibinfo{author}{Kristollari, V.}, \bibinfo{author}{Karathanassi, V.}, \bibinfo{author}{Topouzelis, K.}, \bibinfo{author}{Kolokoussis, P.}, \bibinfo{author}{Taggio, N.}, \bibinfo{author}{Aiello, A.}, \bibinfo{author}{Ceriola, G.}, \bibinfo{author}{Barbone, E.}, \bibinfo{author}{Corradi, P.}, \bibinfo{year}{2022}.
\newblock \bibinfo{title}{Increasing the sentinel-2 potential for marine plastic litter monitoring through image fusion techniques}.
\newblock \bibinfo{journal}{Marine Pollution Bulletin} \bibinfo{volume}{182}.
\bibitem[{Lanaras et~al.(2017)Lanaras, Bioucas-Dias, Baltsavias and Schindler}]{lanaras2017super}
\bibinfo{author}{Lanaras, C.}, \bibinfo{author}{Bioucas-Dias, J.}, \bibinfo{author}{Baltsavias, E.}, \bibinfo{author}{Schindler, K.}, \bibinfo{year}{2017}.
\newblock \bibinfo{title}{Super-resolution of multispectral multiresolution images from a single sensor}, in: \bibinfo{booktitle}{Proceedings of the IEEE Conference on Computer Vision and Pattern Recognition Workshops}, pp. \bibinfo{pages}{20--28}.
\bibitem[{Lanaras et~al.(2018)Lanaras, Bioucas-Dias, Galliani, Baltsavias and Schindler}]{2018-lanaras-network}
\bibinfo{author}{Lanaras, C.}, \bibinfo{author}{Bioucas-Dias, J.}, \bibinfo{author}{Galliani, S.}, \bibinfo{author}{Baltsavias, E.}, \bibinfo{author}{Schindler, K.}, \bibinfo{year}{2018}.
\newblock \bibinfo{title}{Super-resolution of sentinel-2 images: Learning a globally applicable deep neural network}.
\newblock \bibinfo{journal}{ISPRS Journal of Photogrammetry and Remote Sensing} \bibinfo{volume}{146}, \bibinfo{pages}{305--319}.
\bibitem[{Liebel and K{\"o}rner(2016)}]{liebel2016single}
\bibinfo{author}{Liebel, L.}, \bibinfo{author}{K{\"o}rner, M.}, \bibinfo{year}{2016}.
\newblock \bibinfo{title}{Single-image super resolution for multispectral remote sensing data using convolutional neural networks}.
\newblock \bibinfo{journal}{The International Archives of the Photogrammetry, Remote Sensing and Spatial Information Sciences} \bibinfo{volume}{41}, \bibinfo{pages}{883--890}.
\bibitem[{Lim et~al.(2017)Lim, Son, Kim, Nah and Mu~Lee}]{lim2017edsr}
\bibinfo{author}{Lim, B.}, \bibinfo{author}{Son, S.}, \bibinfo{author}{Kim, H.}, \bibinfo{author}{Nah, S.}, \bibinfo{author}{Mu~Lee, K.}, \bibinfo{year}{2017}.
\newblock \bibinfo{title}{Enhanced deep residual networks for single image super-resolution}, in: \bibinfo{booktitle}{Proceedings of the IEEE conference on computer vision and pattern recognition workshops}, pp. \bibinfo{pages}{136--144}.
\bibitem[{Lin and Bioucas-Dias(2020)}]{2020SSSS}
\bibinfo{author}{Lin, C.H.}, \bibinfo{author}{Bioucas-Dias, J.}, \bibinfo{year}{2020}.
\newblock \bibinfo{title}{An explicit and scene-adapted definition of convex self-similarity prior with application to unsupervised sentinel-2 super-resolution}.
\newblock \bibinfo{journal}{IEEE Transactions on Geoscience and Remote Sensing} \bibinfo{volume}{58}, \bibinfo{pages}{3352--3365}.
\bibitem[{Lu et~al.(2023)Lu, Yang, Huang, Chen, Chi, Liu and Tu}]{lu2023awfln}
\bibinfo{author}{Lu, H.}, \bibinfo{author}{Yang, Y.}, \bibinfo{author}{Huang, S.}, \bibinfo{author}{Chen, X.}, \bibinfo{author}{Chi, B.}, \bibinfo{author}{Liu, A.}, \bibinfo{author}{Tu, W.}, \bibinfo{year}{2023}.
\newblock \bibinfo{title}{Awfln: An adaptive weighted feature learning network for pansharpening}.
\newblock \bibinfo{journal}{IEEE Transactions on Geoscience and Remote Sensing} \bibinfo{volume}{61}, \bibinfo{pages}{1--15}.
\bibitem[{Mai et~al.(2024)Mai, Lam and Lee}]{mai2024deep}
\bibinfo{author}{Mai, T.}, \bibinfo{author}{Lam, E.}, \bibinfo{author}{Lee, C.}, \bibinfo{year}{2024}.
\newblock \bibinfo{title}{Deep unfolding tensor rank minimization with generalized detail injection for pansharpening}.
\newblock \bibinfo{journal}{IEEE Transactions on Geoscience and Remote Sensing} .
\bibitem[{Malenovsk{\`y} et~al.(2012)Malenovsk{\`y}, Rott, Cihlar, Schaepman, Garc{\'\i}a-Santos, Fernandes and Berger}]{2012-malenovsky}
\bibinfo{author}{Malenovsk{\`y}, Z.}, \bibinfo{author}{Rott, H.}, \bibinfo{author}{Cihlar, J.}, \bibinfo{author}{Schaepman, M.}, \bibinfo{author}{Garc{\'\i}a-Santos, G.}, \bibinfo{author}{Fernandes, R.}, \bibinfo{author}{Berger, M.}, \bibinfo{year}{2012}.
\newblock \bibinfo{title}{Sentinels for science: Potential of sentinel-1,-2, and-3 missions for scientific observations of ocean, cryosphere, and land}.
\newblock \bibinfo{journal}{Remote Sensing of environment} \bibinfo{volume}{120}, \bibinfo{pages}{91--101}.
\bibitem[{McFeeters(1996)}]{McFEETERS01051996}
\bibinfo{author}{McFeeters, S.K.}, \bibinfo{year}{1996}.
\newblock \bibinfo{title}{The use of the normalized difference water index (ndwi) in the delineation of open water features}.
\newblock \bibinfo{journal}{International Journal of Remote Sensing} \bibinfo{volume}{17}, \bibinfo{pages}{1425--1432}.
\bibitem[{Medoff et~al.(1983)Medoff, Brody, Nassi and Macovski}]{medoff1983iterative}
\bibinfo{author}{Medoff, B.P.}, \bibinfo{author}{Brody, W.R.}, \bibinfo{author}{Nassi, M.}, \bibinfo{author}{Macovski, A.}, \bibinfo{year}{1983}.
\newblock \bibinfo{title}{Iterative convolution backprojection algorithms for image reconstruction from limited data}.
\newblock \bibinfo{journal}{JOSA} \bibinfo{volume}{73}, \bibinfo{pages}{1493--1500}.
\bibitem[{Nguyen et~al.(2021)Nguyen, Ulfarsson, Sveinsson and Dalla~Mura}]{nguyen2021sentinel}
\bibinfo{author}{Nguyen, H.}, \bibinfo{author}{Ulfarsson, M.}, \bibinfo{author}{Sveinsson, J.}, \bibinfo{author}{Dalla~Mura, M.}, \bibinfo{year}{2021}.
\newblock \bibinfo{title}{Sentinel-2 sharpening using a single unsupervised convolutional neural network with mtf-based degradation model}.
\newblock \bibinfo{journal}{IEEE Journal of Selected Topics in Applied Earth Observations and Remote Sensing} \bibinfo{volume}{14}, \bibinfo{pages}{6882--6896}.
\bibitem[{Nguyen et~al.(2023)Nguyen, Ulfarsson, Sveinsson and Mura}]{2023Nguyen}
\bibinfo{author}{Nguyen, H.}, \bibinfo{author}{Ulfarsson, M.}, \bibinfo{author}{Sveinsson, J.}, \bibinfo{author}{Mura, M.}, \bibinfo{year}{2023}.
\newblock \bibinfo{title}{Unsupervised sentinel-2 image fusion using a deep unrolling method}.
\newblock \bibinfo{journal}{IEEE Geoscience and Remote Sensing Letters} \bibinfo{volume}{20}, \bibinfo{pages}{1--5}.
\bibitem[{Palsson et~al.(2018)Palsson, Sveinsson and Ulfarsson}]{ResNet2018}
\bibinfo{author}{Palsson, F.}, \bibinfo{author}{Sveinsson, J.}, \bibinfo{author}{Ulfarsson, M.}, \bibinfo{year}{2018}.
\newblock \bibinfo{title}{Sentinel-2 image fusion using a deep residual network}.
\newblock \bibinfo{journal}{Remote Sensing} \bibinfo{volume}{10}.
\bibitem[{Park et~al.(2017)Park, Choi, Park and Choi}]{2017ModifiedSelSynthBand}
\bibinfo{author}{Park, H.}, \bibinfo{author}{Choi, J.}, \bibinfo{author}{Park, N.}, \bibinfo{author}{Choi, S.}, \bibinfo{year}{2017}.
\newblock \bibinfo{title}{Sharpening the vnir and swir bands of sentinel-2a imagery through modified selected and synthesized band schemes}.
\newblock \bibinfo{journal}{Remote Sensing} \bibinfo{volume}{9}.
\bibitem[{Pereira-S{\'a}nchez et~al.(2024)Pereira-S{\'a}nchez, Sans, Navarro and Duran}]{pereira2024comprehensive}
\bibinfo{author}{Pereira-S{\'a}nchez, I.}, \bibinfo{author}{Sans, E.}, \bibinfo{author}{Navarro, J.}, \bibinfo{author}{Duran, J.}, \bibinfo{year}{2024}.
\newblock \bibinfo{title}{A comprehensive overview of satellite image fusion: From classical model-based to cutting-edge deep learning approaches}.
\newblock \bibinfo{journal}{Super-Resolution for Remote Sensing} , \bibinfo{pages}{279--328}.
\bibitem[{Pereira-Sánchez et~al.(2025)Pereira-Sánchez, Navarro, Petro and Duran}]{2025SSSR}
\bibinfo{author}{Pereira-Sánchez, I.}, \bibinfo{author}{Navarro, J.}, \bibinfo{author}{Petro, A.}, \bibinfo{author}{Duran, J.}, \bibinfo{year}{2025}.
\newblock \bibinfo{title}{Model-guided network with cluster-based operators for spatio-spectral super-resolution}.
\newblock \href{http://arxiv.org/abs/2505.24605}{{\tt arXiv:2505.24605}}.
\bibitem[{Pereira-Sánchez et~al.(2024)Pereira-Sánchez, Sans, Navarro and Duran}]{MARNet}
\bibinfo{author}{Pereira-Sánchez, I.}, \bibinfo{author}{Sans, E.}, \bibinfo{author}{Navarro, J.}, \bibinfo{author}{Duran, J.}, \bibinfo{year}{2024}.
\newblock \bibinfo{title}{Multi-head attention residual unfolded network for model-based pansharpening}.
\newblock \bibinfo{journal}{Submitted} \href{http://arxiv.org/abs/2409.02675}{{\tt arXiv:2409.02675}}.
\bibitem[{Peters(1981)}]{peters1981algorithms}
\bibinfo{author}{Peters, T.M.}, \bibinfo{year}{1981}.
\newblock \bibinfo{title}{Algorithms for fast back- and re-projection in computed tomography}.
\newblock \bibinfo{journal}{IEEE Transactions on Nuclear Science} \bibinfo{volume}{28}, \bibinfo{pages}{3641--3647}.
\bibitem[{Qin et~al.(2024)Qin, Wu, Luo, andD.Dineng Zhao, Zhou, Cui, Wan and Xu}]{2024Bathymetry}
\bibinfo{author}{Qin, X.}, \bibinfo{author}{Wu, Z.}, \bibinfo{author}{Luo, X.}, \bibinfo{author}{andD.Dineng Zhao, J.S.}, \bibinfo{author}{Zhou, J.}, \bibinfo{author}{Cui, J.}, \bibinfo{author}{Wan, H.}, \bibinfo{author}{Xu, G.}, \bibinfo{year}{2024}.
\newblock \bibinfo{title}{Musrfm: Multiple scale resolution fusion based precise and robust satellite derived bathymetry model for island nearshore shallow water regions using sentinel-2 multi-spectral imagery}.
\newblock \bibinfo{journal}{ISPRS Journal of Photogrammetry and Remote Sensing} \bibinfo{volume}{218}, \bibinfo{pages}{150--169}.
\bibitem[{Rabbani and Jones(1991)}]{rabbani1991digital}
\bibinfo{author}{Rabbani, M.}, \bibinfo{author}{Jones, P.}, \bibinfo{year}{1991}.
\newblock \bibinfo{title}{Digital Image Compression Techniques}.
\newblock \bibinfo{edition}{1st} ed., \bibinfo{publisher}{Society of Photo-Optical Instrumentation Engineers (SPIE)}, \bibinfo{address}{USA}.
\bibitem[{Ranchin and Wald(2000)}]{ranchin2000fusion}
\bibinfo{author}{Ranchin, T.}, \bibinfo{author}{Wald, L.}, \bibinfo{year}{2000}.
\newblock \bibinfo{title}{Fusion of high spatial and spectral resolution images: The arsis concept and its implementation}.
\newblock \bibinfo{journal}{Photogrammetric engineering and remote sensing} \bibinfo{volume}{66}, \bibinfo{pages}{49--61}.
\bibitem[{Razzak et~al.(2023)Razzak, Mateo-Garc{\'\i}a, Lecuyer, G{\'o}mez-Chova, Gal and Kalaitzis}]{razzak2023multi}
\bibinfo{author}{Razzak, M.}, \bibinfo{author}{Mateo-Garc{\'\i}a, G.}, \bibinfo{author}{Lecuyer, G.}, \bibinfo{author}{G{\'o}mez-Chova, L.}, \bibinfo{author}{Gal, Y.}, \bibinfo{author}{Kalaitzis, F.}, \bibinfo{year}{2023}.
\newblock \bibinfo{title}{Multi-spectral multi-image super-resolution of sentinel-2 with radiometric consistency losses and its effect on building delineation}.
\newblock \bibinfo{journal}{ISPRS Journal of Photogrammetry and Remote Sensing} \bibinfo{volume}{195}, \bibinfo{pages}{1--13}.
\bibitem[{Salgueiro~Romero et~al.(2020)Salgueiro~Romero, Marcello and Vilaplana}]{salgueiro2020super}
\bibinfo{author}{Salgueiro~Romero, L.}, \bibinfo{author}{Marcello, J.}, \bibinfo{author}{Vilaplana, V.}, \bibinfo{year}{2020}.
\newblock \bibinfo{title}{Super-resolution of sentinel-2 imagery using generative adversarial networks}.
\newblock \bibinfo{journal}{Remote Sensing} \bibinfo{volume}{12}, \bibinfo{pages}{2424}.
\bibitem[{Seager et~al.(2005)Seager, Turner, Schafer and Ford}]{Seager2005VIR}
\bibinfo{author}{Seager, S.}, \bibinfo{author}{Turner, E.}, \bibinfo{author}{Schafer, J.}, \bibinfo{author}{Ford, E.}, \bibinfo{year}{2005}.
\newblock \bibinfo{title}{Vegetation’s red edge: A possible spectroscopic biosignature of extraterrestrial plants}.
\newblock \bibinfo{journal}{Astrobiology} \bibinfo{volume}{5}, \bibinfo{pages}{372–390}.
\bibitem[{Segarra et~al.(2020)Segarra, Buchaillot, Araus and Kefauver}]{2020-agriculture}
\bibinfo{author}{Segarra, J.}, \bibinfo{author}{Buchaillot, M.}, \bibinfo{author}{Araus, J.}, \bibinfo{author}{Kefauver, S.}, \bibinfo{year}{2020}.
\newblock \bibinfo{title}{Remote sensing for precision agriculture: Sentinel-2 improved features and applications}.
\newblock \bibinfo{journal}{Agronomy} \bibinfo{volume}{10}, \bibinfo{pages}{641}.
\bibitem[{Selva et~al.(2015)Selva, Aiazzi, Butera, Chiarantini and Baronti}]{selva2015hypersharpening}
\bibinfo{author}{Selva, M.}, \bibinfo{author}{Aiazzi, B.}, \bibinfo{author}{Butera, F.}, \bibinfo{author}{Chiarantini, L.}, \bibinfo{author}{Baronti, S.}, \bibinfo{year}{2015}.
\newblock \bibinfo{title}{Hyper-sharpening: A first approach on sim-ga data}.
\newblock \bibinfo{journal}{IEEE Journal of selected topics in applied earth observations and remote sensing} \bibinfo{volume}{8}, \bibinfo{pages}{3008--3024}.
\bibitem[{Shettigara(1992)}]{shettigara1992generalized}
\bibinfo{author}{Shettigara, V.}, \bibinfo{year}{1992}.
\newblock \bibinfo{title}{A generalized component substitution technique for spatial enhancement of multispectral images using a higher resolution data set}.
\newblock \bibinfo{journal}{Photogrammetric Engineering and remote sensing} \bibinfo{volume}{58}, \bibinfo{pages}{561--567}.
\bibitem[{Tarasiewicz et~al.(2023)Tarasiewicz, Nalepa, Farrugia, Valentino, Chen, Briffa and Kawulok}]{tarasiewicz2023multitemporal}
\bibinfo{author}{Tarasiewicz, T.}, \bibinfo{author}{Nalepa, J.}, \bibinfo{author}{Farrugia, R.}, \bibinfo{author}{Valentino, G.}, \bibinfo{author}{Chen, M.}, \bibinfo{author}{Briffa, J.}, \bibinfo{author}{Kawulok, M.}, \bibinfo{year}{2023}.
\newblock \bibinfo{title}{Multitemporal and multispectral data fusion for super-resolution of sentinel-2 images}.
\newblock \bibinfo{journal}{IEEE Transactions on Geoscience and Remote Sensing} \bibinfo{volume}{61}, \bibinfo{pages}{1--19}.
\bibitem[{Traganos and Reinartz(2018)}]{2018-seagrass}
\bibinfo{author}{Traganos, D.}, \bibinfo{author}{Reinartz, P.}, \bibinfo{year}{2018}.
\newblock \bibinfo{title}{Mapping mediterranean seagrasses with sentinel-2 imagery}.
\newblock \bibinfo{journal}{Marine pollution bulletin} \bibinfo{volume}{134}, \bibinfo{pages}{197--209}.
\bibitem[{Ulfarsson et~al.(2019)Ulfarsson, Palsson, Dalla~Mura and Sveinsson}]{2019ReducedRank}
\bibinfo{author}{Ulfarsson, M.}, \bibinfo{author}{Palsson, F.}, \bibinfo{author}{Dalla~Mura, M.}, \bibinfo{author}{Sveinsson, J.}, \bibinfo{year}{2019}.
\newblock \bibinfo{title}{Sentinel-2 sharpening using a reduced-rank method}.
\newblock \bibinfo{journal}{IEEE Transactions on Geoscience and Remote Sensing} \bibinfo{volume}{57}, \bibinfo{pages}{6408--6420}.
\bibitem[{Vaqueiro et~al.(2021)Vaqueiro, Fonseca, Oliveira and Mora}]{vaqueiro2021multi}
\bibinfo{author}{Vaqueiro, M.}, \bibinfo{author}{Fonseca, J.}, \bibinfo{author}{Oliveira, H.}, \bibinfo{author}{Mora, A.}, \bibinfo{year}{2021}.
\newblock \bibinfo{title}{Multi-image super-resolution algorithm supported on sentinel-2 satellite images geolocation error}, in: \bibinfo{booktitle}{2021 International Young Engineers Forum (YEF-ECE)}, \bibinfo{organization}{IEEE}. pp. \bibinfo{pages}{50--57}.
\bibitem[{Wald et~al.(1997)Wald, Ranchin and Mangolini}]{wald1997fusion}
\bibinfo{author}{Wald, L.}, \bibinfo{author}{Ranchin, T.}, \bibinfo{author}{Mangolini, M.}, \bibinfo{year}{1997}.
\newblock \bibinfo{title}{Fusion of satellite images of different spatial resolutions: Assessing the quality of resulting images}.
\newblock \bibinfo{journal}{Photogrammetric Engineering and Remote Sensing} \bibinfo{volume}{63}, \bibinfo{pages}{691--699}.
\bibitem[{Wang et~al.(2019)Wang, Huang, Zhang and Ma}]{2019WangSVR}
\bibinfo{author}{Wang, J.}, \bibinfo{author}{Huang, B.}, \bibinfo{author}{Zhang, H.}, \bibinfo{author}{Ma, P.}, \bibinfo{year}{2019}.
\newblock \bibinfo{title}{Sentinel-2a image fusion using a machine learning approach}.
\newblock \bibinfo{journal}{IEEE Transactions on Geoscience and Remote Sensing} \bibinfo{volume}{57}, \bibinfo{pages}{9589--9601}.
\bibitem[{Wang et~al.(2015)Wang, Shi, Atkinson and Pardo-Ig{\'u}zquiza}]{wang2015newgeostatistical}
\bibinfo{author}{Wang, Q.}, \bibinfo{author}{Shi, W.}, \bibinfo{author}{Atkinson, P.}, \bibinfo{author}{Pardo-Ig{\'u}zquiza, E.}, \bibinfo{year}{2015}.
\newblock \bibinfo{title}{A new geostatistical solution to remote sensing image downscaling}.
\newblock \bibinfo{journal}{IEEE transactions on geoscience and remote sensing} \bibinfo{volume}{54}, \bibinfo{pages}{386--396}.
\bibitem[{Wang et~al.(2016)Wang, Shi, Li and Atkinson}]{wang2016}
\bibinfo{author}{Wang, Q.}, \bibinfo{author}{Shi, W.}, \bibinfo{author}{Li, Z.}, \bibinfo{author}{Atkinson, P.}, \bibinfo{year}{2016}.
\newblock \bibinfo{title}{Fusion of sentinel-2 images}.
\newblock \bibinfo{journal}{Remote Sensing of Environment} \bibinfo{volume}{187}, \bibinfo{pages}{241--252}.
\bibitem[{Wang and Bovik(2002)}]{wang2002universal}
\bibinfo{author}{Wang, Z.}, \bibinfo{author}{Bovik, A.}, \bibinfo{year}{2002}.
\newblock \bibinfo{title}{A universal image quality index}.
\newblock \bibinfo{journal}{IEEE Signal Processing Letters} \bibinfo{volume}{9}, \bibinfo{pages}{81--84}.
\bibitem[{Wu et~al.(2023)Wu, Lin, Zhang, Li, Cheng and Nan}]{2023-wu-hyerarchical}
\bibinfo{author}{Wu, J.}, \bibinfo{author}{Lin, L.}, \bibinfo{author}{Zhang, C.}, \bibinfo{author}{Li, T.}, \bibinfo{author}{Cheng, X.}, \bibinfo{author}{Nan, F.}, \bibinfo{year}{2023}.
\newblock \bibinfo{title}{Generating sentinel-2 all-band 10-m data by sharpening 20/60-m bands: A hierarchical fusion network}.
\newblock \bibinfo{journal}{ISPRS Journal of Photogrammetry and Remote Sensing} \bibinfo{volume}{196}, \bibinfo{pages}{16--31}.
\bibitem[{Yan et~al.(2022)Yan, Zhou, Zhang and Xie}]{yan2022mmnet}
\bibinfo{author}{Yan, K.}, \bibinfo{author}{Zhou, M.}, \bibinfo{author}{Zhang, L.}, \bibinfo{author}{Xie, C.}, \bibinfo{year}{2022}.
\newblock \bibinfo{title}{Memory-augmented model-driven network for pansharpening}, in: \bibinfo{editor}{Avidan, S.}, \bibinfo{editor}{Brostow, G.}, \bibinfo{editor}{Ciss{\'e}, M.}, \bibinfo{editor}{Farinella, G.}, \bibinfo{editor}{Hassner, T.} (Eds.), \bibinfo{booktitle}{Computer Vision -- ECCV 2022}, \bibinfo{publisher}{Springer Nature Switzerland}, \bibinfo{address}{Cham}. pp. \bibinfo{pages}{306--322}.
\bibitem[{Yang et~al.(2017)Yang, Fu, Hu, Huang, Ding and Paisley}]{Yang_2017_ICCV}
\bibinfo{author}{Yang, J.}, \bibinfo{author}{Fu, X.}, \bibinfo{author}{Hu, Y.}, \bibinfo{author}{Huang, Y.}, \bibinfo{author}{Ding, X.}, \bibinfo{author}{Paisley, J.}, \bibinfo{year}{2017}.
\newblock \bibinfo{title}{Pannet: A deep network architecture for pan-sharpening}, in: \bibinfo{booktitle}{Proceedings of the IEEE international conference on computer vision}, pp. \bibinfo{pages}{5449--5457}.
\bibitem[{Yokoya et~al.(2011)Yokoya, Yairi and Iwasaki}]{yokoya2011coupled}
\bibinfo{author}{Yokoya, N.}, \bibinfo{author}{Yairi, T.}, \bibinfo{author}{Iwasaki, A.}, \bibinfo{year}{2011}.
\newblock \bibinfo{title}{Coupled nonnegative matrix factorization unmixing for hyperspectral and multispectral data fusion}.
\newblock \bibinfo{journal}{IEEE Transactions on Geoscience and Remote Sensing} \bibinfo{volume}{50}, \bibinfo{pages}{528--537}.
\bibitem[{Zhang et~al.(2024)Zhang, He, Yan, Cao, Li, Xie, Zhou and Hong}]{2024WINet}
\bibinfo{author}{Zhang, J.}, \bibinfo{author}{He, X.}, \bibinfo{author}{Yan, K.}, \bibinfo{author}{Cao, K.}, \bibinfo{author}{Li, R.}, \bibinfo{author}{Xie, C.}, \bibinfo{author}{Zhou, M.}, \bibinfo{author}{Hong, D.}, \bibinfo{year}{2024}.
\newblock \bibinfo{title}{Pan-sharpening with wavelet-enhanced high-frequency information}.
\newblock \bibinfo{journal}{IEEE Transactions on Geoscience and Remote Sensing} \bibinfo{volume}{62}, \bibinfo{pages}{1--14}.
\bibitem[{Zhang et~al.(2023)Zhang, Wang, Zhang, Wan, Sun and Bruzzone}]{zhang2023spatial}
\bibinfo{author}{Zhang, K.}, \bibinfo{author}{Wang, A.}, \bibinfo{author}{Zhang, F.}, \bibinfo{author}{Wan, W.}, \bibinfo{author}{Sun, J.}, \bibinfo{author}{Bruzzone, L.}, \bibinfo{year}{2023}.
\newblock \bibinfo{title}{Spatial-spectral dual back-projection network for pansharpening}.
\newblock \bibinfo{journal}{IEEE Transactions on Geoscience and Remote Sensing} \bibinfo{volume}{61}, \bibinfo{pages}{1--16}.
\bibitem[{Zhang et~al.(2019)Zhang, Su, Fu, Zhu, Xue, Huang, Wang, Wu and Yao}]{zhang2019super}
\bibinfo{author}{Zhang, M.}, \bibinfo{author}{Su, W.}, \bibinfo{author}{Fu, Y.}, \bibinfo{author}{Zhu, D.}, \bibinfo{author}{Xue, J.H.}, \bibinfo{author}{Huang, J.}, \bibinfo{author}{Wang, W.}, \bibinfo{author}{Wu, J.}, \bibinfo{author}{Yao, C.}, \bibinfo{year}{2019}.
\newblock \bibinfo{title}{Super-resolution enhancement of sentinel-2 image for retrieving lai and chlorophyll content of summer corn}.
\newblock \bibinfo{journal}{European Journal of Agronomy} \bibinfo{volume}{111}, \bibinfo{pages}{125938}.
\bibitem[{Zhou et~al.(2022)Zhou, Liu and Wang}]{zhou2022panformer}
\bibinfo{author}{Zhou, H.}, \bibinfo{author}{Liu, Q.}, \bibinfo{author}{Wang, Y.}, \bibinfo{year}{2022}.
\newblock \bibinfo{title}{Panformer: A transformer based model for pan-sharpening}, in: \bibinfo{booktitle}{2022 IEEE International Conference on Multimedia and Expo (ICME)}, \bibinfo{organization}{IEEE}. pp. \bibinfo{pages}{1--6}.
\bibitem[{Zomet et~al.(2001)Zomet, Rav-Acha and Peleg}]{zomet2001robust}
\bibinfo{author}{Zomet, A.}, \bibinfo{author}{Rav-Acha, A.}, \bibinfo{author}{Peleg, S.}, \bibinfo{year}{2001}.
\newblock \bibinfo{title}{Robust super-resolution}, in: \bibinfo{booktitle}{Proceedings of the 2001 IEEE Computer Society Conference on Computer Vision and Pattern Recognition. CVPR 2001}, pp. \bibinfo{pages}{I--I}.

\end{thebibliography}

\end{document}